\documentclass[11pt, a4paper]{article}
\pdfoutput=1
\usepackage[top=2.5cm, bottom=2.5cm,left=2.5cm,right=2.5cm]{geometry}
\usepackage{amsmath,amssymb,graphicx}
\usepackage{tabularx,subcaption,multirow}
\allowdisplaybreaks
\usepackage{cancel,slashed,booktabs}
\usepackage[dvipsnames,table]{xcolor}
\definecolor{myblue}{RGB}{37,52,148}
\definecolor{myred}{RGB}{207,51,73}
\definecolor{mygreen}{RGB}{0,86,79}
\usepackage[utf8]{inputenc}
\usepackage[T1,OT1]{fontenc}
\usepackage{microtype}
\usepackage{appendix}

\usepackage{changes}

\usepackage[style=phys,biblabel=brackets,eprint=true,
    pageranges=false,backend=bibtex,sorting=none]{biblatex}
\newcommand{\gsm}[0]{\mathcal{G}_{\mathrm{SM}}}
\newcommand{\gem}[0]{\mathcal{G}_{\cancel{\mathrm{SM}}}}
\newcommand{\pd}[0]{\textsf{P\textsubscript{D}}}
\newcommand{\sru}[0]{\mathrm{i}}
\newcommand{\half}[0]{\frac{1}{2}}
\newcommand{\ml}{\mathcal{L}}
\newcommand{\LDSF}[0]{\textsf{LDSF}}
\newcommand{\LDSS}[0]{\textsf{LDSS}}
\newcommand{\secant}{\text{sec}}

\DeclareMathOperator{\hc}{\mathrm{h.c.}}

\def \order(#1){{\mathcal O} \left(#1 \right)}

\usepackage{fancyhdr,authblk}
\fancypagestyle{preprint}{%
   \headheight=13.6pt
   \fancyhf{}
   \fancyhead[R]{TIFR/TH/19-39}
}

\usepackage{hyperref}
\hypersetup{%
    linktocpage = true,
    colorlinks = true, 
    urlcolor   = myblue, 
    linkcolor  = mygreen, 
    citecolor  = myblue 
    }
\usepackage[nameinlink]{cleveref}

\title{\LARGE\bf The E\textsubscript{6} route to multicomponent dark matter}
\author[1]{Triparno
Bandyopadhyay\thanks{\href{mailto:triparnb@srmist.edu.in}{triparnb@srmist.edu.in}}}
\author[2,3]{Rinku Maji\thanks{\href{mailto:rinkumaji9792@gmail.com}{rinkumaji9792@gmail.com}}}
\affil[1]{Department of Physics and Nanotechnology, College of
Engineering and Technology,\\ SRM Institute of Science and Technology,
Kattankulathur 603203, Tamil Nadu, India.}
\affil[2]{Department of Physics, Indian Institute of Technology, Kanpur 208016, India.}
\affil[3]{Cosmology, Gravity and Astroparticle Physics Group, Center for
Theoretical Physics of the Universe,\\ Institute for Basic Science, Daejeon 34126, Republic of Korea.}
\date{\today}

\begin{document}
\maketitle
\thispagestyle{preprint}
    \begin{abstract}
        \noindent
        We present a framework of dark- and visible-sector unification in the
        E\textsubscript{6} embedding of the standard model. The demand for
        consistently getting the standard model leads to the existence of the
        dark-sector. We show that the hierarchy of vevs typifying unified models
        leads to multicomponent dark matter at the IR\@. The symmetry breaking
        itself categorises the matter content into dark- and visible-sector
        particles, the categorisation being uniform across different breaking
        chains. We discuss the stability of the dark matter particles and
        compare them to existing phenomenological models of dark matter. The
        central results follow from symmetry and hierarchy arguments. We present
        an indicative set of models of gauge coupling unification, to show that
        the framework can be embedded in realistic models of E\textsubscript{6}.
    \end{abstract}

    \section{Introduction\label{sec:introduction}}%
        With the building blocks of the Standard Model (SM), the
        particles and the interactions, all in place---their properties
        underpinned by experimental data~\cite{Tanabashi:2018oca}---it
        is perhaps time to look at some of the pieces of the overarching
        puzzle that the SM fails to put in place, and to answer some of
        the intriguing questions that its structure begets. The SM, in
        its canonical form, fails to explain neutrino oscillation data~\cite{Esteban2019} and cannot account for the dark matter (DM)
        content of the Universe~\cite{Aghanim:2018eyx}. Models unifying
        the three interactions of the SM~\cite{Georgi:1972cj,
            Georgi:1974my,
        Fritzsch:1974nn} can be ultraviolet (UV) free~\cite{PhysRevLett.30.1346, PhysRevLett.30.1343}, renormalisable~\cite{tHooft:1972tcz}, and protected against quantum gravity
        effects~\cite{Kallosh:1995hi, Hawking1975, Banks:2006mm}, being
        described by a single gauged symmetry at the UV\@. Such
        frameworks predict unification of the gauge couplings and
        partial unification of the Yukawa couplings, giving rise to a
        highly predictive (albeit phenomenological) IR model. Also,
        unified models naturally provide the scales associated with the
        baryon- and lepton-number violating Weinberg operators~\cite{Weinberg:1979sa} related to neutrino see-saw and proton
        decay. Besides, models based on $SO(10)$ and larger symmetries~\cite{Gursey:1975ki,Achiman:1978vg,Shafi:1978gg}
        embed the SM fermions in anomaly free representations, providing
        a non-anthropic explanation for the cancellation of chiral
        anomalies~\cite{Georgi:1980pd} (also check~\cite{Geng:1988pr,
        Minahan:1989vd}).

       If the DM content of the Universe is particle in nature~\cite{Bertone:2004pz, Steigman:2012nb}, then an acceptable model
       of unification should necessarily account for the dark sector, alongside a mechanism to stabilise the DM particle(s). Simple groups of unification with five
        or more diagonal generators can provide a discrete symmetry (\pd) \cite{Kibble:1982ae},
        remnant at the IR, that stabilises the DM particle. The smallest rank-5
        group of unification, $SO(10)$, embeds the fifteen SM fermions and a
        right-handed (RH) neutrino in the spinorial 16-dim representation.
        Hence, there is no place to include DM particles in the same irreducible
        representation (irrep). As for the scalars, all the components of any
        $SO(10)$ irrep, when restricted to the SM gauge symmetry $\gsm=
        SU(3)_c\times SU(2)_w\times U(1)_y$, transform similarly under the
        remnant \pd. Therefore, the scalars responsible for spontaneous symmetry
        breaking (SSB) are necessarily even to keep \pd\ intact. Consequently,
        the irreps containing the SSB scalars cannot include any DM\@. Thus, in
        order to get the dark-sector particles, one has to look for irreps in
        addition to the ones required to get the SM
        \cite{Holman:1982tb, Kadastik:2009cu, Kadastik:2009dj,
        PhysRevD.81.075002, Mambrini:2013iaa, Mambrini:2015vna,
        Nagata:2015dma, Arbelaez:2015ila, Boucenna:2015sdg, Bandyopadhyay:2017uwc,
Ferrari:2018rey}. This introduces much ambiguity in model building,
        destroying the predictability expected from unified models.

        A tidier approach would be to look for unifying frameworks which
        automatically generate the dark-sector while reproducing the
        SM\@. \textcolor{Black}{This unification of the visible- and the
            dark-sectors has been pointed out in Ref.
        \cite{Schwichtenberg:2017xhv}}. The fundamental representation
        of $E_6$, when restricted to $\gsm$, decomposes to different
        sub-multiplets which transform differently under \pd. Therefore,
        the irreps that contain the SM particles can themselves contain
        the dark-sector particles. When embedded in $E_6$, both the SM
        fermions and scalars transform as the 27 dimensional
        fundamental. The additional fermions and scalars in the
        fundamental irrep can then be DM candidates. Therefore, there is
        no arbitrariness in the choice of DM multiplets. In addition,
        alongside unification of the gauge symmetries, the dark- and the
        visible-sector particles are unified. The author
            in Ref. \cite{Schwichtenberg:2017xhv} derives the spectrum
            for the \(E_6\to SO(10)\times U(1)\) symmetry breaking
            route, while in this paper, we study all the symmetry
            breaking routes of $E_6$ down to the SM to determine the
            inherent symmetry structure that simultaneously gives rise
            to the visible- and the dark-sector, irrespective of the
        route of the descent of \(E_6\) down to the SM.


        The results in this paper follow from the following simple
        demands:  i) $E_6$ faithfully reproduces the SM,  ii) the
        symmetry breaking from the unification scale down to the SM
        scale preserves the DM stabilising \pd\ symmetry,  iii) the tiny
        neutrino masses are due to the see-saw
        mechanism, and  iv) a proton decay lifetime that is safe from
        current bounds.  We show that appropriate hierarchies of the
        symmetry breaking scales in the framework leads to a DM sector
        that can simultaneously accommodate two different dark matter
        candidates, with the lightest dark-sector fermion (\LDSF) and
        the lightest dark-sector scalar (\LDSS) being DM candidates,
        both of which can be at the electroweak (EW) scale. The lighter
        of the two is stable while the heavier is metastable. The
        interactions between the two are suppressed by the GUT scale and
        the first generation Yukawa. However, we note that since in a
        unified theory all scalars are expected to be at the highest
        available scale, we have to resort to tuning to keep the mass of
        the lightest dark sector scalar light.
        The heavier component, decays fast
            enough so as to constitute only a tiny fraction of the relic
            density of the dark matter. However, if the mass difference
            is less than the \(W^\pm\) mass, both can have
        \(\mathcal{O}(1)\) relic. The known masses of the SM fermions
        determine the Yukawa couplings of the DM candidates. The results
        discussed do not depend on any particular route of symmetry
        breaking and are valid for all the maximal subgroups of $E_6$,
        namely, $SU(6)\times SU(2)$, $SO(10)\times U(1)$, and
        $SU(3)\times SU(3)\times SU(3)$. Our discussions in this paper
        are restricted to tree-level analyses only and we do not
        consider loop corrections to Fermion masses nor do we analyse
        the Coleman-Weinberg potential \cite{Coleman:1973jx} of the
        model.

        Decaying and multicomponent dark matter \cite{Zurek:2008qg,
        Ibarra:2013cra, Arvanitaki:2008hq} have been studied in the context of
        multiple astrophysical and cosmological observations. For example, in
        the context of relieving the Hubble tension \cite{Vattis:2019efj}, and
        in the context of the CALET and DAMPE excesses \cite{Geng:2019vwm}. In
        this work, our attempt is not to present a new kind of dark matter. On
        the contrary, we are interested in the general structure of SSB that
        stabilises the DM(s) and the qualitative features of the dark-sector. In
        the following sections, we present our analyses. In the next section, we
        look at all the different ways in which the SM hypercharge can be
        constructed out of the diagonal generators of $E_6$ while preserving the
        non-abelian part of $\gsm$. The different definitions readily show which
        of them do preserve \pd\ at the IR and which don't. We show that for the
        cases which can accommodate \pd\ at the IR, the particle content
        transforms identically under $\gsm\times U(1)_D\times U(1)'$, where
        $U(1)_D$ and $U(1)'$ are the additional ranks in $E_6$. It is $U(1)_D$
        that distinguishes between the SM and the dark-sector particles. In
        \cref{sec:DM} we discuss the spectrum and the interactions of the
        dark-sector particles. We also show how two distinct dark matter
        candidates emerge in this framework. We describe the nature of the dark
        matter candidates, both the scalar and the fermion, and point to
        pre-existing analyses which study phenomenological Lagrangians
        containing such DM candidates.  In \cref{sec:guts}, we show gauge
        coupling unification (GCU) for a few symmetry breaking chains to
        establish that our framework can easily be embedded in minimal models of
        $E_6$ unification. Finally, we conclude.

    \section{\texorpdfstring{The E\textsubscript{6} fundamental and a remnant
    Z\textsubscript{2}}{The E6 fundamental and a remnant
    Z2} \label{sec:symmbrk}}%
        As indicated by its name, $E_6$ has six diagonal generators (rank), two
        more than the four diagonals of the SM gauge symmetry, $\gsm\equiv
        SU(3)_c\times SU(2)_w\times U(1)_y$. The additional diagonal generators
        spontaneously break when scalars neutral under $\gsm$, but charged under
        the $U(1)$ symmetries corresponding to the said diagonals, acquire
        non-zero vacuum expectation values (vev). Three of the six diagonal
        generators of $E_6$ give rise to the 2+1 diagonal generators of the
        non-abelian part of the SM ($SU(3)_c\times SU(2)_w$), while a linear
        combination of the three remaining diagonals gives $U(1)_y$. The two
        orthogonal directions are broken in-between the scale of gauge coupling
        unification (GCU) and the scale of electroweak symmetry breaking (EWSB).
        Let $X=(x_1,x_2,x_3)$ be the $U(1)$ charges in the basis where the
        hypercharge is not yet identified. To get the SM hypercharges, we
        perform a rotation on the $X$ basis to get the $Y=(y,y',y'')$ basis,
        $Y_i= O_{x_i x_j}X_j$, where $y$ is the hypercharge and $O_{x_i x_j}$ is
        the rotation matrix. We determine the direction of $y$ from electroweak
        charge assignments.

        \begin{figure}[htpb]
            \centering
            \includegraphics[width=0.7\linewidth]{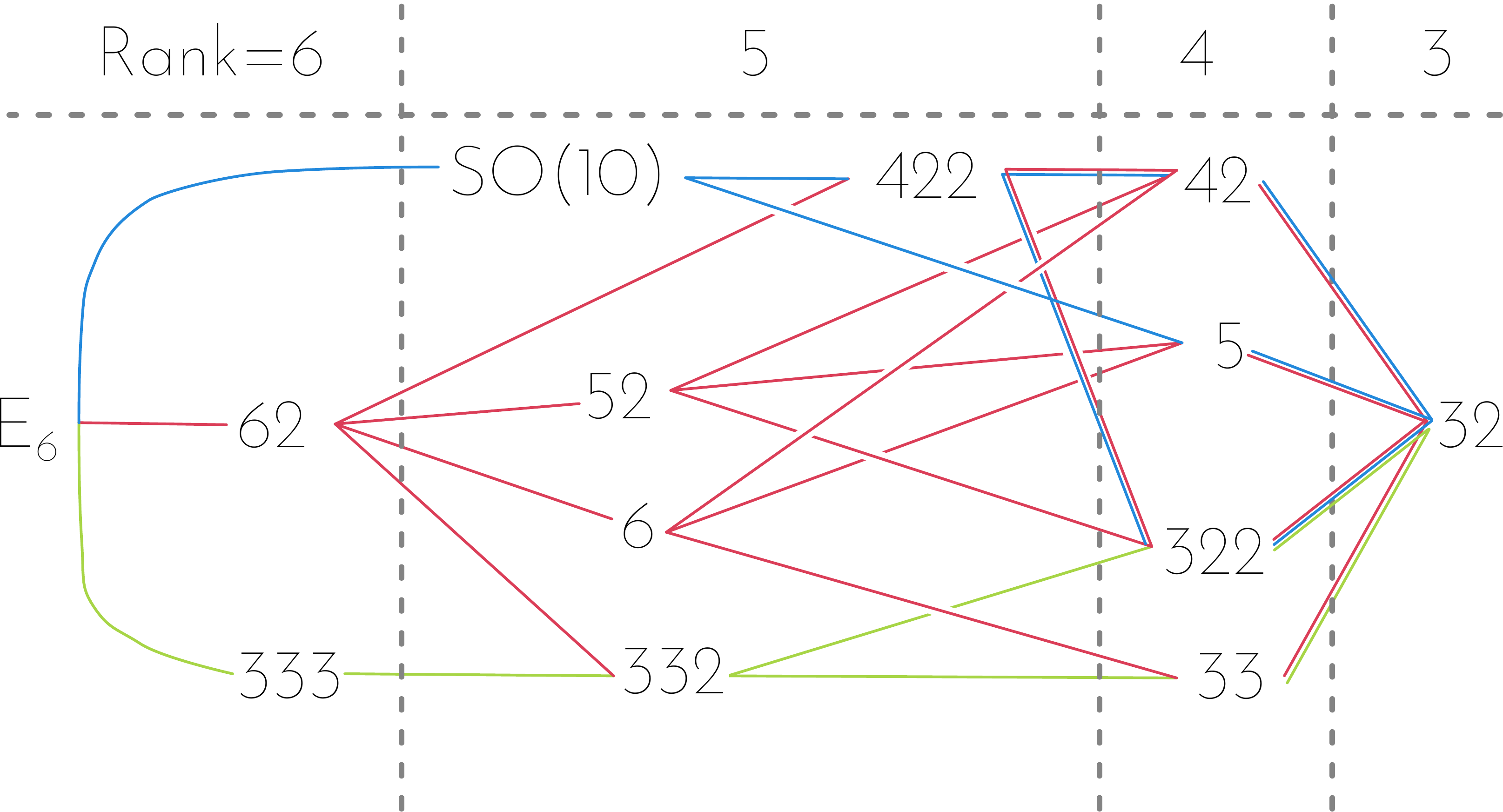}
            \caption{\textit{\small
                Symmetry breaking routes of $E_6$.  We show only the
                non-abelian parts and drop the $U(1)$s. However, we
                indicate the number of broken generators by giving the
                ranks at each stage. We use a compact notation for the
                direct products of the $SU(N)$ symmetries, with
                \textsf{MN} representing $SU(M)\times SU(N)$.
                }
                \label{fig:routes}}
        \end{figure}

        Our goal in this section is to see which symmetry breaking routes of
        $E_6$ contain the DM-stabilising symmetry, \pd. In \cref{fig:routes}, we
        schematically depict the different routes of breaking of $E_6$ down to
        the SM. One can build a plethora of models based on these routes by
        introducing different scalar multiplets to break different symmetries at
        different scales. In the figure, we have only shown the non-abelian
        symmetries, dropping the $U(1)$s for brevity. The ranks in the figure
        indicate the stages at which the abelian symmetries break. The presence
        or absence of a DM stabilising symmetry, in general, depends on the
        model in question. However, as we see below, without going into the
        details of the specific models, we can identify a few symmetry breaking
        chains which cannot, under any circumstance, accommodate a DM
        stabilising symmetry.

        The cyclic group of order $N$, $Z_N$, is a subgroup of $U(1) \forall N$
        ($Z_1$ being the identity). A particle that transforms non-trivially
        under the $U(1)$ will transform under a $Z_N$ symmetry after the $U(1)$
        breaks spontaneously. The `$N$' depends on the charge of the particle
        and that of the $U(1)$ breaking scalar. Let the continuous symmetry be
        broken at some scale $\cancel{M}$\footnote{We will use the `cancelled'
        $M$ notation to denote symmetry breaking scales throughout this paper.}
        by the non-zero vev, $\langle\phi\rangle$, of a scalar $\phi$
        transforming with a charge $x_\phi$. Below $\cancel{M}$, $\psi$- which
        had a charge $x_\psi$ under the $U(1)$ will transform under a remnant
        $Z_N$, defined by \cite{Krauss:1988zc, Ibanez:1991hv, Petersen:2009ip}:
        \begin{align}
            \label{eq:m_par}
            \psi\to \psi'=e^{2\pi\sru\frac{x_\psi}{x_\phi}}\psi.
        \end{align}
        Now, all the scalars in the spectrum that transform under the $U(1)$ and
        acquire a non-zero vevs break it spontaneously. The lowest SSB scale of
        our problem being the EWSB scale ($\cancel{M}_\mathrm{EW}$), the charges
        (if any) of the SM Higgs doublets govern the $Z_N$ transformation of the
        spectrum below $\cancel{M}_\mathrm{EW}$. If the charge of the Higgs is
        such that it breaks the symmetry to identity (following \cref{eq:m_par})
        then, irrespective of the symmetry breaking sequence at the UV, no DM
        stabilising symmetry survives after $\cancel{M}_\mathrm{EW}$. We will
        use this fact along with our knowledge that the SM particle content,
        scalars and fermions, reside in the 27 dimensional $E_6$ fundamental, to
        separate the chains which are suitable for DM model building from the
        ones which are not.

        The $E_6$ fundamental (or anti-fundamental, see \cref{sec:E6}), on restriction
        to $\gsm$, is given by:
        \begin{align}
            \label{eq:res_sm}
            27&\to
             (3,2,1/6)
             + (\overline{3},1,-2/3)
             + (1,1,1)
             + (1,2,1/2)
             + (3,1,-1/3)
             \notag\\&\quad
             + 2\times (\overline{3},1,1/3)
             + 2\times (1,2,-1/2)
             + 2\times (1,1,0).
        \end{align}
        For the case of Weyl fermions (left-handed), we immediately identify $q\equiv
        (3,2,1/6)$, $u^c\equiv (\overline{3},1,-2/3)$, and $e^c\equiv (1,1,1)$. We can
        also identify $(3,1,-1/3)$ and $(1,2,1/2)$ as beyond the Standard Model (BSM) multiplets. There is a
        two-fold ambiguity regarding the down quark $d^c\equiv (\overline{3},1,1/3)$,
        the lepton doublet $\ell\equiv (1,2,-1/2)$, and the singlet neutrino
        $\nu^c\equiv (1,1,0)$. On the other hand, when the 27 contains complex scalars,
        we can unambiguously identify the doublet that couples to the up-type quark,
        $h_u\equiv (1,2,1/2)$. Again, there is a two-fold ambiguity regarding the
        identity of $h_d\equiv (1,2,-1/2)$. The fundamental of $E_6$ is a complex
        representation with $27$ and $\overline{27}$ transforming differently. The only
        $E_6$ invariant involving three 27 dimensional representations is $27\times
        27\times 27$ (not $27\times \overline{27}\times 27$). Therefore, the doublet
        that gives masses to up-type quarks cannot give masses to the down-types, and we
        need two separate doublets, $h_u$ and $h_d$, for the up- and down-type quarks
        respectively. The other multiplets all represent BSM scalars.

        There are three major \emph{routes} for the descent of $E_6$ down to $\gsm$, one
        through $SO(10)\times U(1)$, one through $SU(6)\times SU(2)$, and another
        through $SU(3)\times SU(3)\times SU(3)$. These are the maximal subgroups of
        $E_6$\footnote{There is a fourth maximal subgroup, $F_4$. Unlike the
        others, $F_4$ doesn't reproduce $\gsm$ and is ignored here.}
        and these further break down to $SU(3)\times SU(2)\times U(1)^3$ through
        different intermediate symmetries, like $SU(5)$, $SU(4)\times SU(2)\times
        SU(2)$, $SU(3)\times SU(2)\times SU(2)$ etc., as shown in \cref{fig:routes}. For
        each of the symmetry breaking routes, three different linear combinations of the
        three $U(1)$ symmetries are possible for the hypercharge. Take the $SU(3)\times
        SU(3)\times SU(3)$ case, for example. We identify one of the three $SU(3)$
        symmetries as the $SU(3)_c$ of the SM. One of the two other $SU(3)$ symmetries,
        $SU(3)_w$, is broken to $SU(2)_w$ of the SM. A linear combination of the broken
        diagonal generator, $U(1)_w$, and both the broken diagonals of the third $SU(3)$
        ($SU(3)_N$), gives $U(1)_y$ (See row 1 of \cref{tab:tab1} for details). Two
        possible combinations and the corresponding orthogonal directions are ($T_i$ are
        the generators of $SU(3)$):
        \begin{align}
        \label{eq:eq2a1}
            y&= \frac{1}{6} T^L_8 + \frac{1}{6} T^N_8 \pm T^N_3;\;
            y'= T^L_8 - T^N_8;\; y''= T^L_8 + T^N_8 \mp\frac{1}{3} T^N_3.
        \end{align}
        For this definition, the restriction of the 27 to $\gsm\times U(1)_{y'}\times
        U(1)_{y''}$ is:
        \begin{align}
            \label{eq:nn}
            27&\supset
                \left(3,2,\frac{1}{6}\right)(1,1)
                + \left(\overline{3},1,-\frac{2}{3}\right)\left(1,-\frac{5}{6}\right)
                + (1,1,1)\left(1,\frac{17}{6}\right)
                +(1,1,0)\left(1,\frac{19}{6}\right)
             \notag\\&\quad
                + \left(3,1,-\frac{1}{3}\right)(-2,-2)
                + \left(\overline{3},1,\frac{1}{3}\right)\left(1,-\frac{7}{6}\right)
                +\left(\overline{3},1,\frac{1}{3}\right)(-2,2)
                + (1,1,0)(4,0)
             \notag\\&\quad
                + \left(1,2^*,\pm \frac{1}{2}\right)\left(-2,\mp\frac{1}{6}\right)
                + \left(1,2^*,-\frac{1}{2}\right)(1,-3).
        \end{align}
        We identify $h_u\equiv (1,2^*,1/2)(-2,-1/6)$. We have two options for $h_d$,
        $h_d\equiv (1,2^*,-1/2)(-2,1/6)$ or $h_d\equiv (1,2^*,-1/2)(1,-3)$. For the
        second identification, the $y'$ charges of $h_u$ and $h_d$ are not equal, the
        former being greater than the latter. Therefore, when $h_u$ acquires a non-zero
        vev, $\langle h_u \rangle$, that breaks  $y'$, $h_d$ transforms non-trivially
        under the remnant $Z_N$, according to \cref{eq:m_par}. Hence, the vev of $h_d$,
        $\langle h_d\rangle$, breaks the $Z_N$. If the charge of $h_d$ was greater than
        that of $h_u$, then $\langle h_u\rangle$ would have broken any discrete remnant
        left by $\langle h_d\rangle$. On the other hand, if $h_d$ is identified to be
        $(1,2^*,-1/2)(-2,1/6)$, then $\langle h_d\rangle$ does not break this $Z_2$.  We
        see from \cref{eq:nn}, there are multiplets whose $y'$ charges are coprimes with
        that of $h_u (h_d)$\footnote{`Coprimes' are generally used to describe
        \emph{integers} and not fractions. However, as charges are defined up to a
        multiplicative constant, we can always multiply the $y'$ charges to make them
        all integers. The arguments would remain essentially the same. Hence, we use the
        term `coprimes' for our discussion.}. These multiplets will transform
        non-trivially under the remnant $Z_N$ below the EW scale. For $y''$, we see that
        $h_u$ and $h_d$ again have equal charges. However, the charges of the other
        multiplets are all integer multiples of this charge. Hence, no multiplet
        transforms non-trivially under the $Z_N$ (see \cref{eq:m_par}). It is
        straightforward to see that the $Z_N$ remnant of $y'$ is $Z_2$. In general, if
        the $h_u$, $h_d$ pair have non-zero $y'$ charges then the non-zero vevs of both
        will keep a $Z_N$ unbroken \emph{iff} the charges are equal. The multiplets of
        the spectrum below the symmetry breaking scale will transform under a remnant
        $Z_N$ if the $y'$ charges of those multiplets are coprimes with that of $h_u$
        and $h_d$. If one of the doublets is neutral under $U(1)_{y'}$ then it plays no
        role in the determination of the $Z_N$.

        With $h_{u/d}\equiv (1,2^*,1/2)(-2,\pm1/6)$, at $\cancel{M}_\mathrm{EW}$, the
        sub-multiplets transforms as:
        \begin{align}
            \label{eq:mm}
            27&\supset
            \left(3,2,\frac{1}{6}\right)^-
            + \left(\overline{3},1,-\frac{2}{3}\right)^-
            + \left(\overline{3},1,\frac{1}{3}\right)^-
            + (\overline{3},1,\frac{1}{3})^+
            + \left(3,1,-\frac{1}{3}\right)^+
            \notag\\&
            + \left(1,2^*,\pm \frac{1}{2}\right)^+
            + \left(1,2^*,-\frac{1}{2}\right)^-
            + (1,1,1)^-
            + (1,1,0)^++(1,1,0)^-,
        \end{align}
        where the superscripts $+$ and $-$ denote $Z_2$-even and $Z_2$-odd respectively.
        The other possible definition of hypercharge in this route and the corresponding
        orthogonal directions are given by:
        \begin{align}
            \label{eq:eq2a2}
            y&= \frac{1}{6} T^L_8 - \frac{1}{3} T^N_8;\;
            y_2= T^N_3;\;y_3= T^L_8 + \half T^N_8.
        \end{align}
        For this case, hypercharge is given by a linear combination of the $T_8^L$ and
        $T_8^N$ from $SU(3)_L$ and $SU(3)_N$ respectively, the third broken generator
        being a `spectator'. The restriction of the 27 to $\gsm\times U(1)_{y'}\times
        U(1)_{y''}$ is:
        \begin{align}
            \label{eq:kk}
            27&\to \left(3,2,\frac{1}{6}\right)(0,1)
            + \left(\overline{3},1,-\frac{2}{3}\right)(0,1)
            + \left(3,1,-\frac{1}{3}\right)(0,-2)
            + (1,1,1)(0,1)
            + (1,1,0)\left(\pm\half,\frac{5}{2}\right)
            \notag\\&\quad
            + \left(\overline{3},1,\frac{1}{3}\right)\left(\pm
                \half,-\half\right)
            + \left(1,2^*,-\half\right)\left(\pm\half,-\half\right)
            + \left(1,2^*,\half\right)(0,-2).
        \end{align}
        We immediately see that although $h_u$ preserves a $Z_2$ subgroup
        corresponding to $y''$, $h_d$, irrespective of the identification,
        breaks it down to identity. For $y'$, $h_u$, a singlet, does not break
        it, while $h_d$ breaks it down to identity. Therefore, for this linear
        combination of hypercharge, a DM stabilising symmetry is not present at
        the IR, irrespective of the UV dynamics.

        Performing the same exercise for all the symmetry breaking routes is
        repetitive. Hence, we push it to \cref{sec:E6}. In \cref{tab:tab1}, we
        sequentially present the decomposition of the $E_6$ fundamental under
        $SU(3)_c\times SU(2)_w\times U(1)_{x_1}\times U(1)_{x_2}\times
        U(1)_{x_3}$ for all the different symmetry breaking routes, and mention
        all the possible linear combinations, ($C_1, C_2, C_3$), of the $X$
        charges that give the hypercharge. In \cref{tab:tab2}, we have written
        down the multiplets after the rotation to the $\gsm\times
        U(1)_{y'}\times U(1)_{y''}$ basis. With the preceding discussion, we can
        use the charges in this basis to find out whether a discrete remnant of
        the broken $U(1)$s exist after EWSB or not, as indicated by a
        $\checkmark$ and a $\times$ respectively.  From \cref{tab:tab2}, we see
        that for the cases that preserve a discrete remnant, the spectrum
        transforms identically under $\gsm\times U(1)_{y'}\times U(1)_{y''}$, as
        given in \cref{tab:part}.

        \begin{table}[htpb]
          \centering
        \extrarowheight=\aboverulesep
        \addtolength{\extrarowheight}{\belowrulesep}
          \begin{tabular}{r r c l c c r r l l}
            \toprule
            \multicolumn{3}{c}{Fermions} & $\gsm$ & $U\!(1)'$ & $U\!(1)_\mathrm{D}$ & $\mathrm{P}_\mathrm{D}$ & \multicolumn{2}{c}{Scalars} \\
            \midrule
              && q & $(3,2,\frac{1}{6})$ & -$\frac{\kappa}{3}$ & 1 & $-1$ & $\widetilde{q}$ &&\\
              && $u^c$ & $(\overline{3},1,$-$\frac{2}{3})$ & $\frac{\kappa}{3}$-1 & 1 & $-1$ & $\widetilde{u}^c$ &&\\
              &\multirow{-3}{*}{\rotatebox[origin=b]{90}{Quarks}}
                  & $d^c$ & $(\overline{3},1,\frac{1}{3})$ & $\frac{\kappa}{3}$+1 & 1 &  $-1$ & $\widetilde{d}^c$ &
                  \multirow{-3}{*}{\rotatebox[origin=c]{-90}{Squarks}}&\\
              \cmidrule(l){2-9}
              && $\ell$ & $(1,2$,-$\frac{1}{2})$ & $\kappa$ & 1 & $-1$ & $\widetilde{\ell}$ &&\\
              && $e^c$ & $(1,1,1)$ & 1-$\kappa$ & 1 & $-1$ & $\widetilde{e}^c$ &&\\
              \multirow{-6}{*}{\rotatebox[origin=c]{90}{Visible Sector}}
                  &\multirow{-3}{*}{\rotatebox[origin=c]{90}{Leptons}}
                  & $\nu^c$ & $(1,0,0)$ & -1-$\kappa$ & 1 &  $-1$ & $\widetilde{\nu}^c$ &
                  \multirow{-3}{*}{\rotatebox[origin=c]{-90}{Sleptons}}&
                  \multirow{-6}{*}{\rotatebox[origin=c]{-90}{Dark-sector}}
                  \\
              \midrule
              && $\widetilde{h}_u$ & $(1,2,\frac{1}{2})$ & 1 & -2 & $1$ & $h_u$&&\\
              && $\widetilde{h}_d$ & $(1,2,$-$\frac{1}{2})$ & -1 & -2 & $1$ & $h_d$ &&\\
              &\multirow{-3}{*}{\rotatebox[origin=c]{90}{Higgsinos \rule{0pt}{3pt}}}
                  & $\widetilde{s}_4$ & $(1,1,0)$ & 0 & 4 &  $1$ & $s_4$ &
                  \multirow{-3}{*}{\rotatebox[origin=c]{-90}{Higgses}}&\\
              \cmidrule(l){2-9}
              && $\widetilde{T}$ \rule{0pt}{2em}& $(\overline{3},1$,$\frac{1}{3}$) & $\frac{2\kappa}{3}$ & -2 & 1 & $T$&&\\
              \multirow{-5}{*}{\rotatebox[origin=c]{90}{Dark-sector}}
                  &\multirow{-2}{*}{\rotatebox[origin=c]{90}{Hquarks}}
                  & $\widetilde{T}^c$ \rule[-1em]{0pt}{0pt}& $(3,1$,-$\frac{1}{3})$ & -$\frac{2\kappa}{3}$ & -2& 1 & $T^c$ &
                  \multirow{-2}{*}{\rotatebox[origin=c]{-90}{Lquarks}}&
                  \multirow{-5}{*}{\rotatebox[origin=c]{-90}{Visible Sector}}
                  \\
            \bottomrule
          \end{tabular}
          \caption{\textit{\small We categorise the scalars and the fermions of the fundamental into visible and
                                  dark-sector particles. The multiplets which belong to the dark-sector for the
                                  fermions belong to the visible sector for the scalars, and vice-versa. We further
                                  divide the multiplets into Quarks, Leptons, Higgsinos, and Heavy Quarks for the
                                  Fermions, and Squarks, Sleptons, Higgses, and Leptoquarks for the scalars. For each
                                  of the multiplets we write down the charges under $U(1)_D$ and $U(1)'$, and also the
                                  transformation under \pd. The $-1$ and the $1$ indicate $Z_2$ odd and even respectively.}
        \label{tab:part}}
        \end{table}

        From the seemingly arbitrary charge assignments in \cref{tab:tab2}, we
        obtain the simplified and consistent charge assignments as shown in
        \cref{tab:part} from the simple considerations that the charges are
        defined up to a multiplicative constant and that we can perform a
        rotation in the $(y',y'')$ plane, keeping the SM physics the same.  The
        multiplets have the same charges under $U(1)_{y'}$ across the symmetry
        breaking routes. We identify $U(1)_{y'}$ as $U(1)_D$ (`D' standing for
        dark). Under $U(1)_D$, all of the SM fermions have the same charge
        $x_\mathrm{SM}= 1$, the dark-sector fermions have charge $x_\mathrm{DM}=
        -2$, except for the singlet which has $x_4= 4$. With the Higgs doublets
        carrying a charge $1$ under $U(1)_D$, the SM fermions are odd and the
        dark-sector fermions are even after EWSB. The dark-sector fermions are
        chiral under $U(1)_\mathrm{D}$ and vector-like under $U(1)_{y''}(\equiv
        U(1)')$.  This forces the SM fermions (together with the RH neutrino) to
        cancel all the gauge and the gauge-gravity mixed anomalies among
        themselves for $U(1)'$. Therefore, the $U(1)'$ charges of the SM
        multiplets can be parametrised by one unknown, $\kappa$ (the charge of
        $\ell\equiv (1,2,-1/2)$ for our case), as shown in \cref{tab:part}
        \cite{Bandyopadhyay:2018cwu,Ekstedt:2016wyi}. We then use the
        $U(1)_D\times {U(1)}'^2$ anomaly cancellation conditions to express the
        charges of the dark-sector vector-like triplets with that of the visible
        sector. The value of $\kappa$ is chain dependant. We have given the
        different values of $\kappa$, as obtained from \cref{tab:tab2}, in
        \cref{tab:tab3}\footnote{Therefore, as an added bonus, we get a set of
        anomaly free $Z'$ models from our exercise that can be studied
        separately.}. The key takeaways of this discussion are then:
        \begin{enumerate}
            \item For all the distinct routes of symmetry breaking, corresponding to the
            $SU(6)\times SU(2)$, the $SU(3)\times SU(3)\times SU(3)$, and the
            $SO(10)\times U(1)$ maximal subgroup, there is at least one chain
            (hypercharge definition) for which we get a DM stabilising
            \pd$\equiv Z_2$ at the IR.
            \item For all the chains with \pd, the particle content transforms under
            $\gsm\times U(1)_D\times U(1)'$ exactly the same way, as given in
            \cref{tab:part}. The dark-sector fermions are vector-like under $U(1)'$ and
            hence anomaly free, forcing intra-SM anomaly cancellation. However, under
            $U(1)_D$, the dark-sector fermions are chiral. Hence, for a consistent
            anomaly free theory, the SM and the dark-sector particles need to cancel
            anomaly together, reflecting the fact that the two sectors arise from the
            same multiplet, i.e., dark-visible unification.
            \item The situation is reversed for the scalars. The multiplets which belong
            to the visible-sector for the fermions belong to the dark-sector for the
            scalars and vice-versa.
        \end{enumerate}

        Using these general properties of all the DM stabilising chains of
        $E_6$, we can easily categorise the particles into different sections.
        For the 27 of fermions, the SM particles (and the RH neutrinos) are odd
        under the $Z_2$ symmetry. The exotic fermions are all even and part of
        the dark-sector. As for the 27 of scalars, $h_u$ and $h_d$ are even, and
        the exotics include both odd and even multiplets. The scalars which are
        odd under the $Z_2$ populate a dark-sector while the even ones are part
        of the visible sector. To categorise all the particles of the
        fundamental, we `borrow' nomenclature from supersymmetry, given that for
        each fermion there is a corresponding scalar with same quantum numbers
        (of course, there are three generations of fermions and only one of the
        scalars). There are the SM fermions, the quarks ($q,u^c, d^c$) and the
        leptons ($\ell, e^c,\nu^c$), and corresponding to them there are the
        `squarks' ($\widetilde{q},\widetilde{u}^c, \widetilde{d}^c$) and the
        `sleptons' ($\widetilde{\ell}, \widetilde{e}^c,\widetilde{\nu}^c$). The
        squarks and the sleptons are $Z_2$ odd scalars populating the
        dark-sector, with the neutral components of the sleptons being
        candidates for scalar DM. Then there are the visible sector ($Z_2$ even)
        scalars, the Higgses ($h_u, h_d, s_4$), and corresponding to them, the
        `Higgsinos'($\widetilde{h}_u, \widetilde{h}_d, \widetilde{s}_4$), which
        are part of the dark-sector, the neutral components of which will be
        fermionic DM candidates. Then there are the $Z_2$ even scalar
        leptoquarks ($T,T^c$), and the corresponding exotic coloured fermions,
        which we call the heavy quarks ($\widetilde{T},\widetilde{T}^c$). In
        \cref{tab:part}  we list all the particles along with their labels.

        Till now, we have established the $Z_2$ symmetry remnant after the
        breaking of the additional ranks of $E_6$. We have also derived the
        transformations of the particles under this $Z_2$. We now take a look at
        the possible ramifications of neutrino seesaw on this symmetry. The
        $\nu^c$ are SM singlets and we can write Majorana terms involving them.
        However, under the intermediate symmetries, they typically transform
        non-trivially. Hence, the relevant Majorana mass terms are generated
        from marginal Yukawa terms at the UV. Under the SM gauge symmetry, the
        scalar in that Yukawa will transform as a singlet, say $\delta_R$, and
        below the symmetry breaking scale at which $\delta_R$ gets a vev we
        write the relevant Majorana mass term for the $\nu^c$. The neutrinos
        also get Dirac mass from the doublet $h_u$. The Lagrangian for the
        neutrino masses is:
        \begin{align}
            \label{eq:nss1}
            \ml&\supset -y_D \ell h_u \nu^c - y_M \nu^c \nu^c \delta_R +\hc .
        \end{align}
        The question then is, whether the vev of $\delta_R$ keeps or breaks the
        remnant $Z_2$. Let us consider the $U(1)_D$ symmetry that gets broken to
        the $Z_2$. For the Dirac and Majorana terms to be allowed, we must
        simultaneously have
        \begin{align}
            \label{eq:cc1}
            x_\ell+x_{\nu^c}+x_{h_u}&=0;\; 2x_{\nu^c}+x_{\delta_R}=0.
        \end{align}
        Since the charge of $h_u$ preserves a remnant $Z_2$ under which the SM
        fermions are all odd, we must have $x_\ell= (2n+1)/2 x_{h_u}, n\in
        \mathbb{Z}$, implying $x_{\delta_R}=(2n+3)x_u$. Therefore,
        $x_{\delta_R}$ is an integer multiple of $x_{h_u}$ and hence, by
        \cref{eq:m_par}, any $Z_2$ that is preserved by $h_u$ is also preserved
        by the vev that gives Majorana masses to the neutrinos. Below the scale
        of $\langle \delta_R\rangle$, the lepton doublet\footnote{We use the
        lepton doublet as an example. Any other multiplet can be used for the
        analysis as well.} and the Higgs doublet transforms as
        \begin{align}
          \label{eq:jj}
          \ell\to e^{2\pi\sru\frac{2n+1}{2(2n+3)}}\ell;\; h_u\to e^{\frac{2\pi\sru}{(2n+3)}}h_u.
        \end{align}
        After the breaking of the $U(1)$ by the seesaw scalar, the lepton and
        the SM Higgs are left transforming under a $Z_{2(2n+3)}$ and a
        $Z_{(2n+3)}$ respectively. Now, 2 and $2n+3$ are coprimes $\forall n\in
        \mathbb{Z}$. A cyclic group $Z_{n_1\times n_2}$ can always be decomposed
        to $Z_{n_1}\times Z_{n_2}$ when $n_1$ and $n_2$ are coprimes. Therefore,
        the lepton doublet effectively transforms under a $Z_2\times Z_{(2n+3)}$
        while the  scalar transforms under a $Z_{(2n+3)}$. The vev of the scalar
        then breaks the $Z_{2n+3}$  down to identity leaving the $Z_2$ of the
        lepton doublet intact.  The analysis changes a little if $x_{h_u}=0$,
        but the results remain the same. We discuss such a case in
        \cref{sec:E6}.

         We use the scalar $351'_S$, which contains a $126$
        dimensional $SO(10)$ scalar. The $126$ contains the singlet \(\delta_R\)
        that leads to the Majorana mass term for the neutrinos. In addition, the
        $126$ contains a \(SU(2)\times SU(2)\times SU(4)\) bi-doublet (check
        \cref{tab:irrep}) that we use in addition to the doublets in the
        $10\in27$ to generate realistic SM fermion masses and mixing angles
        \cite{Lazarides:1980nt}. The matter parity of the new doublets coming
        from the bi-doublet is the same as the doublet in the $27$, hence, its
        inclusion does not change the argument presented above.

        In summary, we have categorised all the symmetry breaking chains of
        $E_6$ into those with a DM stabilising symmetry and those without one,
        from the basic demands of reproducing the SM. We found that for all the
        chains that accommodate the $Z_2$, the spectrum transforms in the same
        way. We have also shown that the vev of the scalar that gives Majorana
        mass to the right-handed neutrinos necessarily conserves this $Z_2$.
        There are some additional discussions on this discrete remnant of $E_6$
        in \cref{sec:E6}.  In the next section, we discuss the masses and the
        interactions of the dark-sector particles. We show that under quite
        general circumstances, the dark-sector can accommodate two different
        dark matter particles simultaneously, a real scalar and a Weyl spinor,
        the heavier being meta-stable.

    \section{\textsf{E6DM}, qualitatively\label{sec:DM}}
        In the last section, we categorised the fermions and the scalars in the
        fundamental of $E_6$ into visible and dark-sector particles. We will now
        look at the mass hierarchies and the interactions of the particles given
        in \cref{tab:part}. With $\Psi^I\equiv 27_F, (I=1(1)3)$, $\Phi\equiv
        27_S$, $\Sigma\equiv 351'_S$, the $E_6$ symmetric Lagrangian is (check
        \cref{tab:prod} for the direct products of the 27):
        \begin{align}
            \label{eq:part2}
            \ml&\supset
            {\Psi^I}^\dagger\sru \slashed{D} \Psi^I
            + D_\mu \Phi^\dagger D^\mu \Phi
            + D_\mu \Sigma^\dagger D^\mu \Sigma
            + m_{\Phi}^2 \overline{\Phi}\Phi
            + m_{\Sigma}^2 \overline{\Sigma}\Sigma
            - (Y_{\Phi}^{I\!J}\Psi_I\Psi_{\!J}\Phi
            + Y_{\Sigma}^{I\!J} \Psi_I \Psi_{\! J} \Sigma +\hc)
            \notag\\&\;
            - \lambda_{\Phi} (\overline{\Phi}\Phi)^2
            - \lambda_{\Sigma} (\overline{\Sigma}\Sigma)^2
            - \lambda_{\Phi\Sigma} \overline{\Phi}\Phi \overline{\Sigma}\Sigma
            - (\mu_\Phi \Phi^3 + \mu_\Sigma \Phi^2 \Sigma 
            + \mu_{\Sigma^\prime} {\Sigma}^3
            + \lambda^\prime_{\Phi\Sigma} \Phi \Phi \overline{\Sigma}\overline{\Sigma} + \hc).
        \end{align}
        \begin{subequations}\label{eq:yuk}\noindent
        The $\Sigma$ multiplet contains the SM singlet $\delta_R$ (see
        \cref{eq:nss1}) that gives Majorana masses to the fermions transforming
        as real singlets under $\gsm$ through the $Y_{\Sigma}^{I\!J} \Psi_I
        \Psi_{\! J} \Sigma$ operator. To avoid unnecessary complications, we
        will work with a single generation of fermions and will drop the
        generational indices ($I, J$) in what follows. It is important to
        appreciate that the above Lagrangian is what we need to write to get the
        SM+neutrino mass and no additional operators are present. The fermions
        get Dirac mass from the Yukawa operator $Y_{\Phi}^{I\!J} \Psi_I
        \Psi_{\!J}\Phi$. In terms of the SM sub-multiplets, the Yukawa
        Lagrangian is:
        \begin{align}
            -\ml^Y&= \ml^M_\Phi + \ml^Y_\Phi + \ml^M_\Sigma + \ml^Y_\Sigma,\\
            \mathrm{where,}\notag\\
            \ml^M_\Sigma&=
                  y_\Delta \ell \Delta_L \ell + y_\delta \nu^c \nu^c \delta_R
                  + y_8 \widetilde{s}_4 \widetilde{s}_4 {s}_8
                  + \hc, \label{eq:mmms}
            \\
            \ml^M_\Phi&= \mathcal{L}^M_{SM} + \mathcal{L}^M_{DM},\\
            \ml^Y_\Phi&= \mathcal{L}^Y_{HQ} + \mathcal{L}^Y_{LQ} + \mathcal{L}^Y_{DM},\\
            \mathrm{with,}\notag\\
            \ml^M_{SM}&=
                  y_u q h_u u^c + y_d q h_d d^c
                  + y_\ell \ell h_d e^c + y_\nu \ell h_u \nu^c
                  + y_\Delta \ell \Delta_L \ell + y_\delta \nu^c \nu^c \delta_R
                  + \hc, \\
            \ml^M_{DM}&=
                y_x s_4 \widetilde{h}_u \widetilde{h}_d
                + y_1 \widetilde{s}_4 \widetilde{h}_u {h}_d
                + y_2 \widetilde{s}_4 \widetilde{h}_d {h}_u
                + y_c s_4 \widetilde{T} \widetilde{T}^c
                + \hc, \label{eq:dmmas}
                \\
            \ml^Y_{HQ}&=
                  (y'_1 d^c \widetilde{\nu}^c + y''_1 \widetilde{d}^c {\nu}^c
                  + y'_2 u^c \widetilde{e}^c + y''_2 e^c \widetilde{u}^c
                  + y'_3 q \widetilde{q}) \widetilde{T}^c
            \notag\\&\quad
                  + (y'_4 q \widetilde{\ell} + y''_4 \ell \widetilde{q}
                  + y'_5 d^c \widetilde{u}^c + y''_5 u^c \widetilde{d}^c) \widetilde{T} + \hc, \\
            \ml^Y_{LQ}&= T
                (\widehat{y}_3 d^c  \nu^c
                + \widehat{y}_4 q q
                + \widehat{y}_7 \widetilde{s}_4 \widetilde{T}
                + \widehat{y}_2 u^c e^c
                )
                + T^c(\widehat{y}_1 q \ell
                + \widehat{y}_5 u^c d^c
                + \widehat{y}_6 \widetilde{s}_4 \widetilde{T}^c)
                +\hc, \\
            \ml^Y_{DM}&=
                  \widetilde{h}_u (
                    y_1 \ell \widetilde{\nu}^c + \overline{y}_1 \widetilde{\ell} \nu^c
                    + y_3 q \widetilde{u}^c + \overline{y}_3 \widetilde{q} u^c
                  )
                + \widetilde{h}_d (
                  y_2 \ell \widetilde{e}^c + \overline{y}_2 \widetilde{\ell} e^c
                  + y_4 q \widetilde{d}^c + \overline{y}_4 \widetilde{q} d^c
                ) + \hc.
                \label{eq:smdm}
        \end{align}
        \end{subequations}
        We have written \emph{all} the terms involving the sub-multiplets in
        $\Phi$ and only the two terms that are relevant to fermion masses for
        the sub-multiplets in $\Sigma$. Note, there are some terms which are
        allowed by $\gsm\times Z_2$ at the IR but not by the unifying symmetry
        at the UV (check \cref{tab:prod} for the direct products of the irreps
        of $E_6$ and $SO(10)$). Given that Yukawa couplings are technically
        natural \cite{Hooft1980}, they will not run to non-zero values from the
        zero fixed point at the $E_6$ breaking scale and hence, have been
        ignored. All the Yukawa couplings corresponding to $\Phi$ (similarly for
        $\Sigma$) are the low-energy manifestations of $Y_\Phi$. Therefore, they
        are all expected to be of the same order (especially, given the fact
        that Yukawa couplings self renormalise). In the following discussion we
        will use this multiple times.

        In $\ml^M_\Sigma$, we collect the terms which give Majorana masses to
        the fermions from the non-zero vevs of the $Z_2$ even singlet and
        triplet\footnote{When we refer to multiplets using their dimensions
        without any further qualifications, we always imply the dimensionality
        under $SU(2)_w$. All scalars that have non-zero vevs are $SU(3)_c$
        singlets.} scalars in $\Sigma$:
        \begin{align}
            \label{eq:scal_vev_0}
            s_8&= \frac{v_8 + \phi' + \sru a_8^0}{\sqrt{2}},
            \delta_R= \frac{v_\mathrm{ss} + \delta' + \sru a_R^0}{\sqrt{2}},
            \Delta_L= \begin{pmatrix} \delta_L^+/\sqrt{2} & \delta_L^{++} \\
                      \frac{v_L + \delta_L' + \sru a_L^0}{\sqrt{2}} & -\delta_L^+/\sqrt{2}
                      \end{pmatrix}.
        \end{align}
        The scalars $\delta_R, \Delta_L\in
        (126,2)_{10}/(1,6,\overline{6})_{333}/(105',1)_{62}$ are responsible for
        the Majorana masses of the left- and right-handed neutrinos, as given in
        \cref{eq:nss1}, and result in a neutrino mass-matrix of:
        \begin{align}
          \label{eq:nss2}
             M_\nu&= \begin{pmatrix} y_\Delta v_l & y_\nu v_u \\
                      y_\nu v_u & y_\delta v_\mathrm{ss}
                    \end{pmatrix},
        \end{align}
        where, due to the presence of both right-handed neutrinos and a triplet
        scalar, we have an admixture of type-I and type-II seesaw
        \cite{Minkowski:1977sc,Yanagida:1979as,Mohapatra:1979ia,Schechter:1980gr}.
        The present uncertainty on the SM $\rho$ parameter
        \cite{Tanabashi:2018oca} constraints $v_l$ to be very small, $\lesssim
        1$ GeV  \cite{Veltman:1977kh}, and in this work we take it to be zero.
        $\ml^M_\Phi$ includes the terms which give Dirac masses to the fermions
        from the non-zero vevs of the $Z_2$ even singlet and doublet scalars in
        $\Phi$:
        \begin{align}
            \label{eq:scal_vev}
            h_u&= \left(h_u^+,\frac{v_u+h_u^0+\sru a_u^0}{\sqrt{2}}\right);\,
            h_d=  \left(\frac{v_d+h_d^0+\sru a_d^0}{\sqrt{2}},h_d^-\right);\,
            s_4=  \frac{v_4+h_4^0+\sru a_4^0}{\sqrt{2}}.
        \end{align}
        In $\ml^Y$, we collect the terms involving scalars that do not get a vev, with
        $\ml_{LQ}^Y$ describing interactions among the fermions and the two leptoquarks,
        $T$ and $T^c$. Models where $T$ and $T^c$ couple in a generation non-universal
        way to the SM fermions have recently gathered some attention in the context of
        the $R_{K^{(*)}}$ anomalies \cite{Bauer:2015knc}. However, these leptoquarks, in
        general, mediate proton decay \cite{Weinberg:1979sa, Dorsner:2012nq}, with the
        bound being $\gtrsim 10^{12}$ GeV. The leptoquarks and the squarks
        ($\widetilde{q}, \widetilde{u}^c$, and $\widetilde{d}^c$),  do not play any role
        in the discussions of dark matter properties and we keep them at the unification
        scale, following the extended survival hypothesis
        \cite{Georgi:1979md, delAguila:1980qag, Mohapatra:1982aq}.

        \subsection{Dark-sector Fermions and Scalars}%
        \label{sub:dsm}
        The mass spectrum of the Higgsinos is determined by the charge assignments, the
        Higgsinos (and the heavy quarks) are vector-like under $U(1)'$ and chiral under
        $U(1)_D$. Therefore, the Higgsinos get a mass at the $U(1)_D$ breaking scale
        and, are blind to $U(1)'$ breaking. Also, $U(1)_D$ divides the Higgsinos into
        two species, with $x_D=-2$ and with $x_D=4$. Therefore, the same scalar cannot
        give mass to both the species. The $x_D=-2$ species readily get masses from
        $\langle s_4\rangle$. The SM singlet $\widetilde{s}_4$ remain massless  when
        only the scalars of the fundamental and $\delta_R, \Delta_L$
        (\cref{eq:scal_vev_0,eq:scal_vev}) get non-zero vevs. There is, however, a SM
        singlet scalar in $\Sigma$, $s_8\equiv(1,1,0)(0,8), (\in (1,-8)_{101}/(1,6,\overline{6})_{333}/(15,1)_{62}) $, which has a Yukawa coupling
        with $\widetilde{s}_4$, and can give it mass, as shown in \cref{eq:mmms}. The
        mass spectrum then depends on the relative sizes of the vevs of the two $\gsm$
        singlet scalars,
        \begin{align}
            \label{eq:scal_vev_dm}
            s_4&=  \frac{v_4+h_4^0+\sru a_4^0}{\sqrt{2}}\in \Phi,\,
            s_8= \frac{v_8 + \phi' + \sru a_8^0}{\sqrt{2}}\in \Sigma,
        \end{align}
        relative to $\cancel{M}_\mathrm{EW}$. The neutral-Higgsino mass matrix, as
        obtained from $\ml^M_{DM}+\ml^M_{SS}$ in the basis $(\tilde{s}_4, \tilde{h}_u^0,
        \tilde{h}_d^0)$, is:
        \begin{align}
            \label{eq:Higgsino_diag}
            M_{ \widetilde{h}_0}&= \begin{pmatrix}
             y_8 v_8 & y_1 v_d & y_2 v_u \\
             y_1 v_d & 0 & y_x v_4 \\
             y_2 v_u & y_x v_4 & 0
             \end{pmatrix}
        \end{align}
        The scalar $s_4$, transforming under the 27 of $E_6$, breaks the unifying group
        after acquiring a vev. Therefore, if we use it for the breaking of $E_6\to
        \mathcal{G}_I$, then, with $v_{d/u}\sim \cancel{M}_\mathrm{EW}<<v_4\sim
        \cancel{M}_U$, there is a hierarchical separation between the eigenvalues of the
        above matrix (with $v_8$ at a scale intermediate to the two). The diagonalising
        matrix, $R_\chi$, defined by $(\widetilde{s_4}, \widetilde{h}_u^0,
        \widetilde{h}_d^0)^T= R_\chi (\chi_1, \chi_2, \chi_3)^T$, $\chi_i$ being the
        diagonal basis, can be written as:
        \begin{align}
            \label{eq:chi_rot}
            R_{\chi}&=  \frac{1}{\sqrt{2}}
                     \begin{pmatrix}
                        0 & \sqrt{2} \cos\theta_\chi & -\sqrt{2} \sin\theta_\chi \\
                        1 & \sin\theta_\chi & \cos\theta_\chi \\
                        -1 & \sin\theta_\chi & \cos\theta_\chi \\
                    \end{pmatrix}.
        \end{align}
        Defining the mass matrix in the diagonal basis as $M_\chi=\mathrm{diag}(M_1,
        M_2, M_3)$, we get:
        \begin{subequations}
        \begin{align}
            M_1&= - y_x v_4\equiv -M,\\
            \tan 2\theta_\chi&
            \simeq -2\sqrt{2} \frac{v_u}{v_4}\sim -2\sqrt{2} \frac{\cancel{M}_\mathrm{EW}}{\cancel{M}_U},\label{eq:chiang}\\
                M_2&= y_8 v_8 -M\secant 2\theta_\chi\sin^2\theta_\chi\simeq
                y_8 v_8-2M(v_{EW}/v_U)^2+\mathcal{O}((v_{EW}/v_U)^3),
                \label{eq:dferum_ang}\\
                M_3&= M\secant 2\theta_\chi\cos\theta_\chi^2=
                M(1+\mathcal{O}((v_{EW}/v_U)^2)).
        \end{align}
        \end{subequations}
        As mentioned previously, $y_1, y_2$, and $y_x$ are all low energy manifestations
        of $Y_{27}$, as given in \cref{eq:yuk}, hence, are of the same order. Similarly,
        the Yukawas $y_8$ and $y_\nu$ are both low energy manifestations of the
        $Y_{351}$ and hence are of the same order. The scale of the seesaw vev $v_\nu$
        is fixed by running of the couplings. Then, $y_\nu$ is fixed by the masses of
        the light neutrinos. Therefore, the order of $y_8$ is also fixed, making the
        scenario quite predictive. We can absorb the $-ve$ sign corresponding to $M_1$
        as a phase by redefining the fermion fields.

        The mixing between the doublets and the singlet, \cref{eq:chiang}, is
        unification scale suppressed, with $\sin\theta_\chi\sim y_x
        \cancel{M}_\mathrm{EW}/\cancel{M}_U\simeq 10^{-20}$, where we use $y_x\sim
        10^{-6}$ as this is the same Yukawa that suppresses the electron mass wrt
        $\cancel{M}_\mathrm{EW}$. We then have two superheavy semi-degenerate Weyl
        fermions, $\chi_1$ and $\chi_3$, and a third fermion $\chi_2\equiv \chi$ with
        mass $\sim y_8 v_8$. In terms of the mass eigenstates, the flavour eigenstates
        are:
        \begin{subequations}
            \label{eq:def_hig}
            \begin{align}
                \tilde{s}_4&= \cos\theta_\chi \chi-\sin\theta_\chi \chi_3,\\
                \tilde{h}_{d,u}&= \pm \frac{1}{\sqrt{2}}\chi_1
                        + \frac{1}{\sqrt{2}}\sin\theta_\chi \chi
                        + \frac{1}{\sqrt{2}}\cos\theta_\chi \chi_3.
            \end{align}
        \end{subequations}
        The lightest dark-sector fermion (\LDSF), $\chi$, is almost
        entirely composed of $\tilde{s}_4$, while $\chi_1$ and $\chi_3$
        have equal contributions from $\tilde{h}_u^0$ and
        $\tilde{h}_d^0$. We see from \cref{eq:yuk} that all the marginal
        interactions of $\tilde{s}_4$ with the SM particles involve the
        $\widetilde{h}_{u,d}$. Hence, in the physical basis, are all
        $\sin\theta_\chi$ suppressed. The \LDSF\ does have the Yukawa
        interactions with $\phi$, the scalar eigenstate composed
        predominantly of $s_8$. Therefore, the fermion DM candidate acts
        like a single Majorana DM talking to the SM through a singlet
        scalar\footnote{As a side note, in the limit $v_8\to 0$, the
            \LDSF\ is almost massless and inert. The bulk astrophysical
            properties of ultralight fermionic dark matter,  with almost
            no couplings with the SM and hence treated as a Thomas-Fermi
            fluid has been recently studied \cite{Pal:2019tqq}. The DM
            mass under consideration in that analysis is $\sim$eV, much
            larger than what we get in this scenario. However, in a more
            realistic and intricate model, if the value of the vev
            playing the role of $v_4$ comes down (i.e.
            \(v_4\) is decoupled from \(\cancel{M}_U\)), then $\chi_2$
            can be in the eV range and act like the particle as
            discussed in that analysis.}. The scalar $s_8$ transforms as
            $(1,1,0)(-8,0)$ under $\gsm\times U(1)_D\times U(1)'$.
            Therefore, its vev, and consequently the mass of $\chi$, can
            be kept at any scale below the intermediate symmetry
            breaking scale (i.e. where the gauge symmetry is $\gsm$).
            The mass of $\chi$ is then a quasi-free parameter of the
            theory. Finally, we assume a normal mass
                hierarchy fermions in the dark sector. That is, we take
                the second and third generation Fermions to be heavier
            than the first generation ones, mimicking the SM.

        \begin{figure}[htpb]
          \centering
          \includegraphics{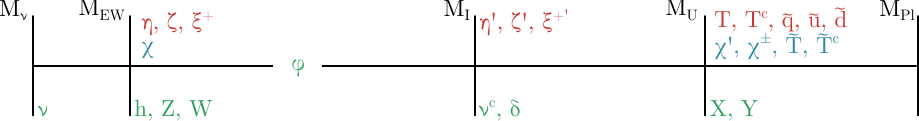}
          \caption{\textit{\small Hierarchy of scales in \textsf{E6DM}.
                  The particles in Red are dark-sector scalars while
                  those in Blue are dark-sector fermions. The ones in
                  green are visible sector particles. In the text, we
                  describe the masses of which particles are associated
                  to symmetry breaking scales, which are fixed using the
                  extended survival hypothesis, and which are
                  (quasi-)free parameters. The scale extends from the
                  masses of the light-neutrinos on the left to the
                  Planck scale on the right. Between these two, are the
                  EWSB scale, the unification scale ($X$ and $Y$ gauge
                  bosons), and the intermediate rank breaking (neutrino
                  seesaw) scale. The state $\phi$, the mass eigenstate
                  corresponding to $s_8$, can vary over a range between
                  the SM and the intermediate scales. The scale of
                  $\phi$ is a parameter controlling the interaction
          strength between the \LDSF\ and the SM Higgs. We describe the
  case where the \LDSF\ and the \LDSS\ are both at the EW scale. Other
possibilities also exist.}}
          \label{fig:Scal}
        \end{figure}
 
            We note that \(s_8\) can develop an induced vev due to its
            trilinear coupling to, \emph{e.g.}, \(s_4\). This coupling
            originates from the \(E_6\) invariant \(\Phi\Phi\Sigma\).
            Once the \(\Sigma\) is integrated out, a dim-5 term,
            \(\Psi\Psi\Phi^*\Phi^*\to \widetilde{s}_4\widetilde{s}_4 s_4
            s_4\) is generated. A simple way out of this is to take the
            trilinear coupling corresponding to \(\Phi\Phi\Sigma\) to be
            very small. This term is expected to push the mass of
            \(\widetilde{s}_4\) to the scale of the \(s_4\) vev \(\sim
            \cancel{M}_U\). As our scalar spectrum contains both a
            \(27\) and a \(351^\prime\), we can have the induced vev
            from the \(27\times 27\times 351^\prime\), or from loops, to
            be cancelled by that from the \({351}^\prime\times
            {351}^\prime\times 351^\prime\) trilinear, as given in
            \cref{eq:part2}. It can be checked from the table in
            \cref{tab:prod} that the \(351^\prime\) contains a singlet
            of \(U(1)_D\) charge 4 that couples with \(s_8\). This
            cancellation amounts to a fine tuning. Such fine tunings,
            albeit undesirable, are necessary to obtain a
            phenomenologically significant spectrum in unified models.
            For example, the fine tuning of the trilinears is similar to
            the one applied in unified models with D-parity broken
            left-right symmetry~\cite{Chang:1983fu}. The addition of
            scalars on top of those required for symmetry breaking is
            quite inevitable in most GUT models, for example, we need
            additional scalar multiplets  to reproduce the correct mass
            spectrum of the SM fermions (\emph{e.g.}
            \cite{Georgi:1979ga, Lazarides:1980nt, Bajc:2005zf,
            Ohlsson:2019sja}).  

        The charged components of the Higgsinos get Dirac mass from the
        same Yukawa as their neutral counterparts and are at the same
        scale. The heavy quarks, $\widetilde{T}, \widetilde{T}^c$, also
        get masses from the same vev as the doublet Higgsinos, as can be
        checked from \cref{eq:dmmas}. Therefore, when the vev of $s_4$
        is at the unification scale, the charged Higgsinos and the heavy
        quarks are also pushed there, along with the neutral components
        of the doublet Higgsinos. This ensures that the heavy quarks are
        not the lightest dark-sector particles, protecting them from the
        stringent bounds from direct detection and IceCube data on
        strongly interacting DM particles \cite{Albuquerque:2003ei,
        Albuquerque:2010bt}. In \cref{fig:Scal}, we schematically depict
        the relevant mass scales in this framework. The visible-sector
        particles are in Green and the dark-sector fermions are in blue.
        The doublet-Higgsinos and the heavy-quarks are depicted at the
        unification scale, to be accurate, they are at the scale where
        $U(1)_D$ breaks. Here, in line of the discussion above, we
        identify it as the unification scale. In the next section, we
        will show a case where there is no rank breaking at the
        unification scale, with $U(1)_D$ breaking coming down to some
        intermediate value. The singlet Higgsino has been kept at the
        electroweak scale, we will discuss more about the choice in the
        next subsection. In Red, we have the dark-sector scalar, the
        masses of which will we discuss now.

        The scalar potential, as given in \cref{eq:part2}, has a global $U(1)$
        preserving part and a $U(1)$ breaking part. The latter forces a mass splitting
        between the real and imaginary parts of the neutral dark scalars, as we show
        below. Also, the cubic and quartic terms involving $\Phi$ and $\Sigma$ ensure
        that all the visible sector scalars, except the SM Higgs, are at the scale of
        $v_\mathrm{ss}$ (the intermediate scale of neutrino seesaw) \cite{Babu:2015bna,
        Chakrabortty:2016wkl}. Hence, at the EW scale, we only need to worry about the
        125 GeV mass eigenstate. In terms of the relevant $\gsm$ sub-multiplets, the
        interactions of the dark-sector scalars, as given in \cref{eq:part2} are:
        \begin{align}
            \label{eq:scal_pot_1}
            V_\mathrm{DM}&=
                       m^2_{\ell\ell} |\widetilde{\ell}|^2
                     + m^2_{\nu\nu} |\widetilde{\nu}|^2
                     + m^2_{ee} |\widetilde{e}|^2
                     - \lambda_\ell |\widetilde{\ell}|^4
                     - \lambda_\nu |\widetilde{\nu}|^4
                     - \lambda_e |\widetilde{e}|^4
            \notag\\&\quad
                     - \lambda_{\ell\nu} \widetilde{\ell}^\dagger \widetilde{\ell}
                        \widetilde{\nu}^* \widetilde{\nu}
                     - \lambda_{\ell e} \widetilde{\ell}^\dagger \widetilde{\ell}
                        \widetilde{e}^* \widetilde{e}
                     - \lambda_{e\nu} \widetilde{e}^* \widetilde{e}\,
                      \widetilde{\nu}^* \widetilde{\nu}
            \notag\\&\quad
                      - (\lambda_{u\ell} |h_u|^2 + \lambda_{d\ell} |h_d|^2 )
                        |\widetilde{\ell}|^2
                       - \lambda_{ux} |h_u^\dagger \widetilde{\ell}|^2
                       - \lambda_{dx} |h_d^\dagger \widetilde{\ell}|^2
                       - \lambda_{x\ell} \left(h_d^\dagger \ell\right)^2
            \notag\\&\quad
                      - ( \lambda_{u\nu} |h_u|^2 + \lambda_{d\nu} |h_d|^2) |\tilde{\nu}|^2
                       - (\lambda_{ue} |h_u|^2 + \lambda_{de} |h_d|^2 ) |\tilde{e}|^2
            \notag\\&\quad
                       - \mu_u h_u  \widetilde{\ell} \widetilde{\nu}
                       - \mu_d h_d \tau_2\widetilde{\ell} \widetilde{e}
                       - \mu_\delta \widetilde{\nu} \widetilde{\nu} \delta
                    + \mathcal{O}\!\left(\frac{1}{M_\mathrm{ss}}\right)
                    + \hc.
        \end{align}
        The contributions to the scalar masses from the intermediate and the unification
        scales, through quartics like $\overline{\Phi}\Phi\overline{\Sigma}\Sigma$ etc.,
        are included in the quadratic couplings, $m_{ii}$. The global $U(1)$ of the
        complex scalars are broken by the $\lambda_{(u/d)x}$ quartics and the $\mu_i$
        cubics. This results in the splitting of the masses between the CP-odd and
        CP-even neutral scalars. Then, with
        \begin{align}
            \label{eq:scal_brk}
            \widetilde{\ell}\ &=\ \begin{pmatrix}
                \left(\widetilde{\ell}_0^r+i\widetilde{\ell}_0^i\right)/\sqrt{2}\\
                        \widetilde{\ell}_\pm 
            \end{pmatrix}\;,\quad \widetilde{\nu}\ =\
        \frac{\widetilde{\nu}^r+i\widetilde{\nu}^i}{\sqrt{2}}\;,
        \end{align}
        the squared-mass matrix, $M^2_S$($M^2_P)$, for the
        CP-even(odd) scalars in the flavour basis $(\widetilde{\ell}_0^r,
        \widetilde{\nu}^r)$ ($(\widetilde{\ell}_0^i, \widetilde{\nu}^i)$) is given by:
        \begin{subequations}
            \label{eq:mscal_ev}
        \begin{align}
            M^2_S&=
             \frac{1}{2} \begin{pmatrix}
                2 m^2_{\ell\ell} - v_u^2 \lambda_{u\ell} - v_d^2 (
                \lambda_{d\ell} + \lambda_{dx} + 2 \lambda_{x\ell})
                        & \sqrt{2} v_u \mu_u \\  \sqrt{2} v_u \mu_u &
                2 m^2_{\nu\nu} - v_d^2 \lambda_{d\nu} - v_u^2
                \lambda_{u\nu} - 2 \sqrt{2} v_R \mu_\delta
                  \end{pmatrix}\\
            M^2_P&=
            \frac{1}{2} \begin{pmatrix}
                2 m^2_{\ell\ell} - v_u^2 \lambda_{u\ell} - v_d^2 (
                \lambda_{d\ell} + \lambda_{dx} - 2 \lambda_{x\ell})
                        & -\sqrt{2} v_u \mu_u \\  -\sqrt{2} v_u \mu_u &
                2 m^2_{\nu\nu} - v_d^2 \lambda_{d\nu} - v_u^2
                \lambda_{u\nu} + 2 \sqrt{2} v_R \mu_\delta
                  \end{pmatrix}.
        \end{align}
        \end{subequations}
        The opposite signed off-diagonal elements in the two matrices are due to the
        difference in phase. The $U(1)$ breaking by the trilinear term corresponds to
        the lepton number breaking Majorana neutrino mass term in the fermion
        sector\footnote{Indeed, as the $\nu^c$ fermion and $\widetilde{\nu}$ scalar
        transform the same way under the symmetries of the Lagrangian, it is not a
        surprise that the global $U(1)$ corresponding to number conservation for both
        them would be the same. Hence, the same scalar that breaks the $U(1)$ for
        $\nu^c$, breaks the same for $\widetilde{\nu}$. The quartic $U(1)$ breaking
        spurions, $\lambda_{(u/d)x}$, have no analogue in the fermionic sector at the
        renormalisable level but correspond to the Lepton number violating dim-5
        Weinberg operators of the lepton doublets}. The trilinear and quartic $U(1)$
        breaking spurions then introduce a mass difference between the CP-even and
        odd scalars, as can be seen from \cref{eq:mscal_ev}.

        The quartic, $\lambda_{(u/d)x}$, spurions also causes mass
        splitting between the charged and neutral components of
        $\widetilde{\ell}$. However, the charged scalar masses do not
        have contributions from the $U(1)$ breaking trilinear, ensuring
        the absence of charge breaking vacua.  The charged scalar mass
        matrix in the \((\tilde{\ell}_+^i, \tilde{e}^c)\) basis is found to be:
          \begin{equation}
            \label{eq:csm1}
            M^2_\pm=
            \frac{1}{2} \begin{pmatrix}
                2 m^2_{\ell\ell} - v_d^2 \lambda_{d\ell} - v_u^2 (\lambda_{u\ell} + \lambda_{ux})
                        & -\sqrt{2} v_d \mu_d \\  -\sqrt{2} v_d \mu_d &
                2 m^2_{ee} - v_d^2 \lambda_{de} - v_u^2 \lambda_{ue}
                  \end{pmatrix}.
        \end{equation}
        Therefore, if all the quartics of $\Phi$ with other scalars,
        like $\Sigma$ and the $E_6$ breaking scalar, are tuned to be
        zero, then the charged scalars will be at the EW scale.
        Nevertheless, these couplings will be generated at higher
        orders, as quartic couplings are not self renormalising, to push
        the scalar masses to the intermediate or the unification scale.
        In general, to get scalars, dark or visible, at the EW scale we
        need to fine-tune the parameters of the scalar sector. We see
            from \cref{eq:mscal_ev,eq:csm1} that there are multiple
            parameters which can be tuned to do so\footnote{This is in
                principle an expansion of the scope of the extended
                survival hypothesis \cite{Georgi:1979md,
delAguila:1980qag, Mohapatra:1982aq} for scalars. }.  We can make one
CP-even, one CP-odd scalar and one charged scalar light with minimal
tuning. They are mostly comprised of the doublet. In that case,
defining the mass basis for the CP-even scalars as,
        \begin{align}
            \label{eq:scal_def}
            \begin{pmatrix}\eta \\ \eta' \end{pmatrix}&=
              \begin{pmatrix} \cos\phi_0 & \sin\phi_0 \\
                -\sin\phi_0 & \cos\phi_0
              \end{pmatrix}
              \begin{pmatrix}
                \widetilde{\ell}_0^r \\ \widetilde{\nu}^c_r
              \end{pmatrix},
        \end{align}
        the mixing angle is given by
        \begin{align}
          \label{eq:scal_mix}
           \sin2\phi_0&= \frac{\sqrt{2}v_u\mu_u}{M^2_{\eta'}-M^2_\eta}
           \implies \sin\phi_0\simeq
           \frac{\sqrt{2}v_\mathrm{EW}\mu_u}{M^2_\mathrm{SS}}.
        \end{align}
        Now, $\mu_u$ is a remnant of the $\Phi\Phi\Sigma$ trilinear given in
        \cref{eq:part2}. There is no `natural' scale for the mass-dimensional
        coefficient of the trilinear terms. Hence, it is a free parameter of the
        Lagrangian. In general, we see from the above relation that the mixing is
        intermediate scale suppressed.

        Similarly, defining
        \begin{align}
            \label{eq:scal_oth_def}
            \begin{pmatrix}
                \zeta \\ \zeta^\prime
            \end{pmatrix}\ &=\ 
            \begin{pmatrix} 
                \cos\phi_p & \sin\phi_p \\
                -\sin\phi_p & \cos\phi_p 
            \end{pmatrix} 
            \begin{pmatrix} 
                \tilde{\ell}_0^i \\
                \tilde{\nu}^i
            \end{pmatrix} \;,\quad
            \begin{pmatrix}
                \xi^+ \\ \xi^{+^\prime}
            \end{pmatrix}\ =\ 
            \begin{pmatrix} 
                \cos\phi_+ & \sin\phi_+ \\
                -\sin\phi_+ & \cos\phi_+
            \end{pmatrix} 
            \begin{pmatrix} 
                \tilde{\ell}_+ \\
                \tilde{e}^c
            \end{pmatrix}\;, 
        \end{align}
        and with \(m_{ee}\) and \(m_{\nu\nu}\) at the intermediate
        scale, we have,
        \begin{align}
            \label{eq:ang2}
            \sin\phi_p\ &\simeq\
            -\frac{\sqrt{2}v_\mathrm{EW}\mu_u}{M^2_\mathrm{SS}}\,;\quad
            \sin\phi_+\ \simeq\
            -\frac{\sqrt{2}v_\mathrm{EW}\mu_d}{M^2_\mathrm{SS}}\,.
        \end{align}
        That is, both these mixings are intermediate scale suppressed as
        well.

        The key takeaways from the scalar sector are that the CP-even
        scalars, the CP-odd scalars, and the charged scalar eigenvectors
        are not aligned and hence there is mass splitting at the tree
        level itself. Both the lightest CP-odd and CP-even scalar could
        be the DM candidate. We choose to work with the CP-even scalar
        as the \LDSS, but the other choice is as valid.
        We use the mass splitting discussed above to
            keep the doublet-like scalars, CP-odd, CP-even, and singly
            charged, at the EW scale, and the predominantly singlet ones
            at the intermediate scale, in a way similar to what is done
            for the visible sector particles \cite{Babu:2015bna,
            Chakrabortty:2016wkl, Georgi:1979md, delAguila:1980qag,
        Mohapatra:1982aq}. We use the mass splitting among the
        components of the doublet to make the CP-even scalar the
    lightest of the three. For these \emph{choices} of the
    charged-scalar and the pseudo-scalar masses, the \LDSS\ is a
    real-singlet at the IR\@. It annihilates to the SM through its
    couplings with the SM Higgs, as we discuss below. In
    \cref{fig:Scal}, we diagrammatically show the hierarchy in the
    scalar sector. The leptoquarks and the squarks are kept at the
    unification scale, as mentioned above. The doublet-like CP-even
    neutral scalar, $\eta$, CP-odd neutral scalar \(\zeta\), and the
    charged scalar \(\xi^+\) are at the EW scale, while all the other
sleptons are at the intermediate scale.

        It is needless to say that an analysis of a complete
        loop-corrected potential \emph{\`a la} Coleman-Weinberg is
        necessary to properly establish the legitimacy of the fine
        tunings which we use for the tree level Lagrangian and to check
        whether more scalar multiplets are necessary to impose all the
        tunings. However, such an analysis is beyond the scope or the
        intent of this paper and we keep it for a future endeavour.

        \subsection{Decays and annihilations}%
        \label{sub:dna}
        At tree level, the \LDSF\ ($\chi$) and the \LDSS\ ($\eta$)
        interact through extremely feeble interactions generated from
        the Yukawa terms in \cref{eq:yuk}, to be precise:
        \begin{subequations}
        \begin{align}
            \label{eq:smdm2}
             \ml\supset
                 y'_1 {\ell \widetilde{h}_u \widetilde{\nu}^c}
                + y'_2 {\ell \widetilde{h}_d \widetilde{e}^c}
                + y_4 e^c \widetilde{\ell} \widetilde{h}_d
                + y_5 \nu^c \widetilde{\ell} \widetilde{h}_u,
        \end{align}
        which in the mass basis is written as:
        \begin{align}
            \label{eq:smdm21}
             \ml\supset
             \frac{\sin\theta_\chi}{\sqrt{2}} \left(
             {y'_1}  \sin\phi_0 \eta \nu
                + y_5 \sin\alpha \cos\phi_0 \nu  \eta
            \right)\chi,
        \end{align}
        \end{subequations}
        where, $\sin\alpha\sim \sqrt{M_\nu/M_{\nu^c}}$
        \cite{delAguila:2008cj} is the mixing between the left- and
        right-handed neutrinos. Therefore, at tree-level, all two-body
        decays of the \LDSS\ to \LDSF\ or vice-versa are
        $\sin\theta_\chi$, $\sin\phi_0$\footnote{The $\sin\alpha$
        suppressed channel is even smaller for $M_\nu\sim$ eV. Hence, we
    have ignored it here.}, and $y'_1(y_5)\sim y_e$ suppressed. Now,
    $\sin\theta_\chi\sim \frac{v_\mathrm{EW}}{M_U}$
    (\cref{eq:dferum_ang}) and $\sin\phi_0\sim
    \frac{v_\mathrm{EW}\mu_u}{M^2_{\nu^c}}$ (\cref{eq:scal_mix}). This
    implies,
        \begin{align}
            \label{eq:}
            \Gamma^\mathrm{tree}_{\eta\to \chi\nu}\sim 2 y_e^2 v^4_\mathrm{EW}\frac{1}{M^2_U}\frac{\mu_u^2}{M^4_\mathrm{\nu^c}}
            M_\eta,
        \end{align}
        where we choose $M_\eta>M_\chi$ (the other choice is also valid). Therefore,
        there is a double suppression of scales, the unification scale and the seesaw
        scale.  
           \begin{figure}[htpb]
           \centering
           \includegraphics{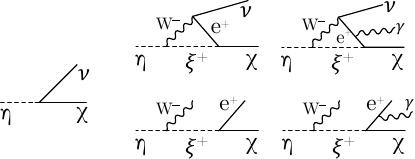}
           \caption{\textit{On the left we show the two body decay mode of the
           \LDSS\ at tree level. On the top-row at right, we show the two one-loop
           diagrams leading to the decay of the \LDSS\ to a neutrino (two-body) and to a
           neutrino and a photon (three-body). On the right bottom-row  we show the
           $\xi^+$ mediated tree-level decay to a neutrino (three-body) and to a
           neutrino and a photon (four-body).}} 
           \label{fig:decay}
         \end{figure}
 {\color{Black}

     One-loop and multibody decays of the \LDSS\ to the \LDSF\ is
     generated by its interactions with the charges sleptons. In the
     mass basis, obtained from \cref{eq:smdm2}:
        \begin{align}
            \label{eq:smdm22}
             \ml\supset
             \frac{\sin\theta_\chi}{\sqrt{2}} \left(
              {y'_2} \sin\phi_+\xi^+ e
             + y_4 \cos\phi_+ \xi^- e^c
            \right)\chi,
        \end{align}
        In \cref{fig:decay}, we show some of these decay modes. Decays
        involving the first term are \(\cancel{M}_U\) and intermediate
        scale suppressed as before, however, those mediated by the
        second term are only \(\cancel{M}_U\) suppressed. However, the
        lack of intermediate scale suppression is compensated by \(G_F\)
        and chiral suppressions. The loop induced decay width is 
        \begin{align}
            \label{eq:loop}
            \Gamma^\mathrm{loop}_{\eta\to \chi\nu}\sim
            \frac{G_F^2}{16\pi^2} m_e^4
            \frac{v_\mathrm{EW}^2}{M_U^2} M_\eta\;,
        \end{align}
        with \(M_W, M_\eta, M_\xi\sim \cancel{M}_\mathrm{EW}\). The
        electron mass in the numerator is due to the vev insertion in
        the electron propagator and due to the \(\mathcal{O}(y_e)\)
        Yukawa coupling of the \(e^c\) with \(\xi^+\) and \(\chi\).
        Taking $v_\mathrm{EW}\sim 10^2$ GeV, $m_e\sim 10^{-4}$ GeV,
        $\cancel{M}_U\sim 10^{16}$ GeV, we have, for both tree and loop,
        $\tau_{\eta\to \chi\nu}\sim 10^{32}$ s. This lifetime is much
        larger than the age of the Universe $\sim 10^{17}$s and also
        much larger than the bounds set on decaying dark dark matter,
        $\tau> 10^{24}$ s for $M_\eta\sim 100$ GeV, set by indirect
        observations of the neutrino spectra \cite{Covi:2009xn,
        Ibarra:2013cra}. These analyses derive the lower limit on the
        lifetime of the decaying dark matter particle  using data on
        atmospheric neutrino flux as obtained from the Frejus
        \cite{Frejus:1994brq}, the AMANDA-II \cite{IceCube:2009ckd}, and
        the Super-Kamiokande \cite{Super-Kamiokande:2005mbp}
        collaborations, along with decaying dark matter searches by the
        IceCube collaboration \cite{IceCube:2011kcp}, and also Refs. 
        \cite{Honda:2006qj,Gonzalez-Garcia:2006xta}. Although we take
        generic values for the scales to calculate the decay width, in
        the next section, we will derive the scales in a few models from
        the demand of GCU.

        More stringent limits on decaying dark matter come from the
        observations of the isotropic gamma ray background
        \cite{Blanco:2018esa, Mambrini:2015sia, Cirelli:2012ut} and the
        cosmic microwave background (CMB) \cite{Slatyer:2016qyl}. These
        limits severely bound scenarios with
        $\tau_{\eta\to\gamma}<10^{28}$s for $M_\eta\sim
        \cancel{M}_\mathrm{EW}$. As we show in \cref{fig:decay}, in our
        setup, the decay of the DM to a photon is a three-body process
        at one-loop and a four-body process at tree-level. Moreover, it
        decays to charged SM particles via (tree-level) three and
        four-body decay channels. These decay channels provide the
        dominant contribution to the \LDSS\ decay width and pose the most
        stringent constraint on its contribution to the relic density.
        We impose a conservative bound $\tau_\eta\gtrsim 10^{28}$ s for
        $\Omega_\eta/\Omega_\chi=1$, where $\Omega$ denotes the relic
        density ($\Omega_{\rm DM}=\Omega_\eta+\Omega_\chi$). The
        numerical values of the partial decay widths are dependant on
        the mass of the $\xi^+$ ($\sim M_\eta$).
        If we take the mass difference between the \LDSS\ and \LDSF\
        ($\Delta_D$) to be greater than $M_W$, so that the SM $W$ bosons
        are produced on-shell. The partial lifetimes, $\tau_{\eta\to
        W^-e^+\chi}$, for three body decay at tree level, 
        is depicted in \cref{fig:3body_dec}.
        \begin{figure}[htbp]
        \begin{center}
        \includegraphics[width=0.45\textwidth]{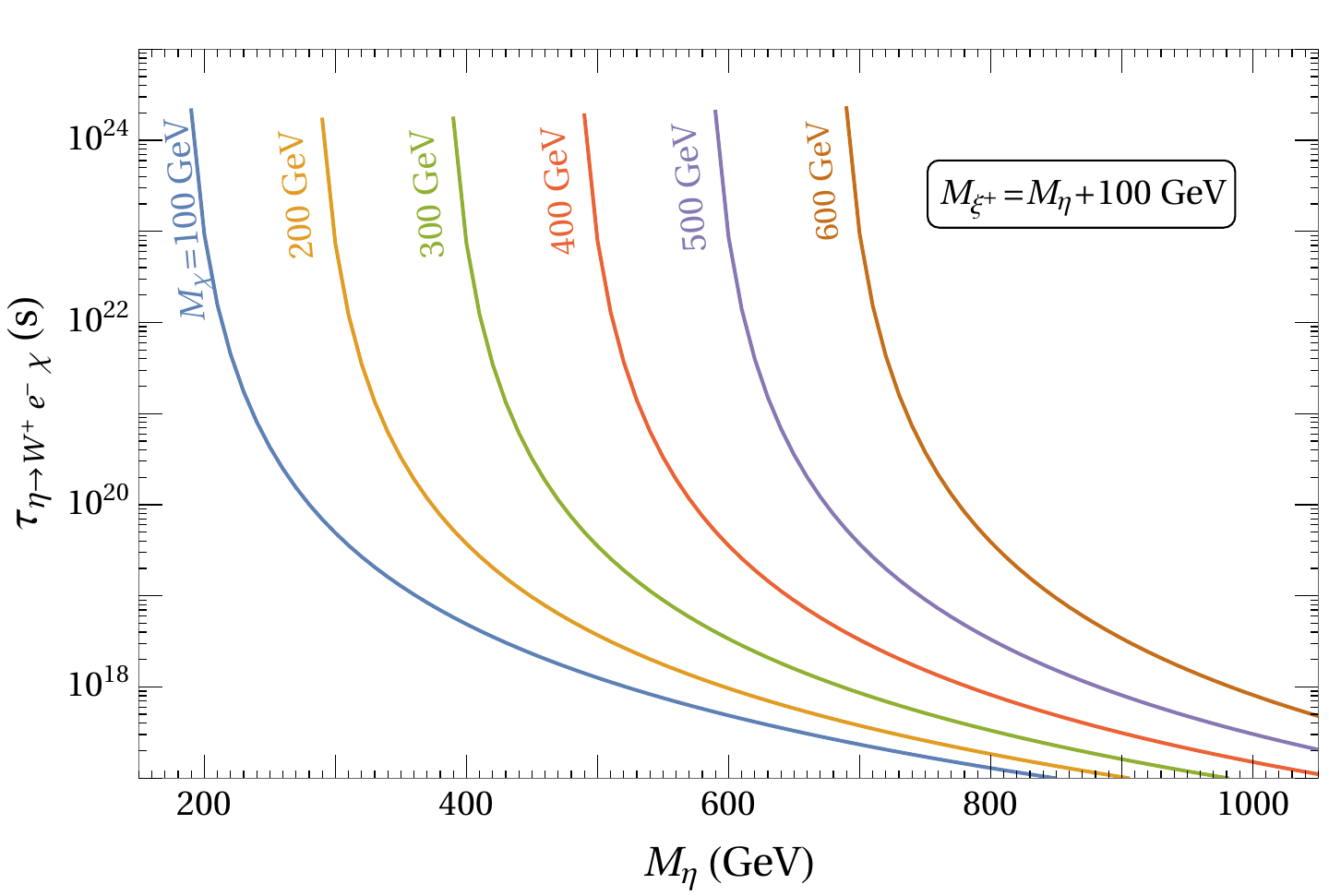}
        \includegraphics[width=0.45\textwidth]{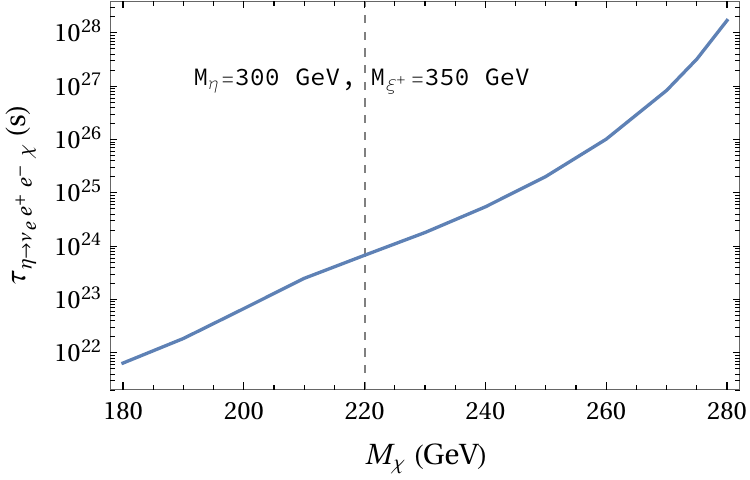}
        \end{center}
        \caption{\textit{\small Left: Partial decay lifetime for the
                channel $\eta\to \chi W^+ e^-$. We have plotted the
                lifetime as a function of $M_\eta$ for the different
                values of $M_\chi$ and the charged slepton mass is set
                to a value $M_{\xi^+}=(M_\eta + 100~\mathrm{GeV})$.
            Right: Partial decay lifetime for the channel $\eta\to \chi
    \nu_e e^+ e^-$. We have plotted the lifetime as a function of $M_\chi$
    for a typical choice $M_\eta=300$ GeV and the charged slepton mass
    $M_{\xi^+}=350$ GeV using \textsf{CalcHEP}
    \cite{Belyaev:2012qa}.}}
          \label{fig:3body_dec}
        \end{figure}
        We can see $10^{24}\gtrsim \tau_{\eta\to W^-e^+\chi}\gtrsim
        10^{18}$~s for $\cancel{M}_\mathrm{EW}<M_\eta<1\;\mathrm{TeV}$.
        Therefore, $\eta$ can comprise upto a tiny fraction
        ($\Omega_\eta \lesssim \frac{\tau_{\eta\to
        W^-e^+\chi}}{10^{28}~\mathrm{s}}\Omega_{\rm DM}$) of the dark
        matter. On contrary, if $\Delta_D<M_W$, the tree-level four-body
        decay will be mediated by an offshell $W$-boson. The four-body
        decay will be further suppressed and the partial lifetime will
        be $\tau \gtrsim 10^{24}$ s (see the right panel of
        \cref{fig:3body_dec} as an example). We then get a scenario
        where the metastable component is   fraction,
        $\Omega_\eta/\Omega_\chi\gtrsim 10^{-4}$, of the total relic.
        
        One of the two DM candidates is then absolutely stable, by
    virtue of the DM stabilising $Z_2$, \pd. The other one is metastable
and is constrained to be only a minute fraction by the current bounds on
decaying dark matter.}  Here, we have assumed the \LDSS\ to be heavier,
however, if the \LDSF\ and the \LDSS\ are interchanged, the conclusions
remain the same as the amplitudes are the same (up to the half factor
coming from averaging of incoming fermion spin). At the IR ,the DM
particles act like, depending on the mass-splitting, one
    real-singlet scalar and one real-singlet fermion, annihilating to
    the SM through the Higgs boson, \emph{or} one real-singlet fermion
and an inert scalar doublet. Such phenomenological models of DM have
been extensively studied in the literature and, for the sake of
completeness, we summarise some of these results. For a
recent study of multicomponent dark matter model that consists of an inert
Higgs doublet and a fermionic component see \cite{Betancur:2020fdl}.
       
        Unification scale suppressed dimension-six operators leading to viable decaying
        dark matter candidates have been studied in the literature before
        \cite{Arvanitaki:2008hq, Ibarra:2013cra}. In such models, the decay width is
        suppressed by $1/\cancel{M}_U^4$. In our case, the suppression is
        $1/\cancel{M}_U^2$, with further suppression coming from the seesaw scale and
        the electron Yukawa. At this point, we note that it is crucial for the CP-even
        scalar coming from the doublet, $\widetilde{\ell}$, to be the \LDSS. If the \LDSS\
        is dominated by $\widetilde{\nu}^c$, then the operator $y_1' \ell\widetilde{h}_u
        \widetilde{\nu}^c$ will lead to $y_1'\sin\theta_\chi\cos\phi_0\eta\nu/\sqrt{2}$
        in the mass basis. In this case, there is no $\sin\phi_0$ suppression in the
        amplitude, and the DM lifetime will not be large enough to avoid existing
        bounds.  The two-component DM paradigm does not work in that case.

        The \LDSF, $\chi$, interacts with the SM through $\cancel{M}_U$
        and $M_{\nu^c}$ suppressed higher dimensional operators. However,
        it does have a Yukawa interaction with the singlet $s_8$, as
        given in \cref{eq:mmms}. The quartic coupling of $s_8$ with the
        SM Higgs doublets introduces mass mixing between the real part
        of $s_8$, $\phi'$ (\cref{eq:scal_vev_dm}), and the real part of
        the scalars in the doublets. At leading order, this mixing is
        \begin{align} 
            \sin\beta\sim \frac{1}{y_8}\frac{M_h M_\chi}{M_\phi^2}, 
        \end{align} 
        where $M_\phi$ is the mass of
        $\phi$, the mass eigenstate corresponding to the real part of
        $s_8$, and we have used $M_\chi\sim y_8 v_8$ as per our
        derivation in \cref{eq:dferum_ang}. This mixing induces a Yukawa
        coupling between $\chi$ and the SM Higgs, $h$. There is also a
        dim-5 operator which is generated due to the $s_8^* s_8
        h^\dagger_{u/d} h_{u/d}$ quartics, after one $s_8$ has been
        replaced by the vev and the other has been integrated out. We
        parametrise the interactions as: \begin{align} \label{eq:ldsfc}
        \ml&= - C'_\chi \frac{v_\mathrm{EW}M_\chi}{M_\phi^2} \chi\chi h
    - C_\chi \frac{M_\chi}{M_\phi^2} \chi \chi h h, \end{align} where we
    have replaced $v_8$ with $M_\chi$ in the second term and absorbed
    the $\mathcal{O}$(1) $y_8$ into $C^{(\prime)}_\chi$. A detailed
    phenomenological study of singlet Majorana dark matter can be found
    in \cite{Baek:2011aa, Matsumoto:2014rxa,
    Matsumoto:2016hbs,GAMBIT:2018eea}. From the analyses in these
    papers, we find that a Majorana singlet, $\chi$, with 1 GeV
    $\lesssim M_\chi \lesssim$ 1 TeV, satisfy the requirement of the
    thermal relic $\Omega_c \leq 0.12 h^2$ ($h$ is the reduced Hubble
    constant) \cite{Aghanim:2018eyx} for a large range of coupling
    strengths as calculated in the usual framework \cite{Bertone:2004pz,
        Bertone:2010zza, Steigman:2012nb}\footnote{Two of the analyses
        mentioned above, viz. \cite{Baek:2011aa, Matsumoto:2014rxa}, use
        a slightly different value for the thermal relic of cold dark
        matter $\Omega_c \lesssim 0.1 h^{-2`}$, based on older
        measurements by the WMAP and the Planck collaboration. However,
    as this is an upper limit, the slightly lower value used in these
analyses does not rule out their conclusions.}. Global fits
\cite{Liu:2017kmx} of the parameters of the effective Lagrangian, using
direct-detection data from the XENON100 \cite{XENON100:2012itz}, LUX
\cite{LUX:2016ggv}, XENON1T \cite{XENON:2017vdw}, and PandaX
\cite{PandaX-II:2016vec} collaborations, also concur (Check
\cite{Dutra:2015vca,Arhrib:2018eex} for similar models of singlet
Majorana DM.).

        The constraints on the trilinear coupling is much more severe
        than the quartic, as it leads to DM-nucleon scattering at
        leading order. In \cite{Arcadi:2019lka}, the authors show that
        taking into account the measurements from the XENON1T experiment
        \cite{XENON:2017vdw}, the net coupling is constrained to be
        $\lesssim 0.01$ for $10$ GeV $<M_\chi<1$ TeV. For the trilinear
        coupling in \cref{eq:ldsfc}, this implies $M_\phi>1$ TeV for an
        $\mathcal{O}$(1) $C'_\chi$ and $M_\chi\sim
        \cancel{M}_\mathrm{EW}$. The bounds on the coupling of the
        quartic are much more relaxed because it leads to DM-nucleon
        interactions only at one-loop, as can be seen from
        \cite{Liu:2017kmx}. As an endnote to this discussion, we note
        that the tiny, high-scale-suppressed, couplings of the \LDSF\ to
        the SM particles might also motivate a freeze-in like scenario
        where the \LDSF\ is produced from the scattering of two SM
        particles. Freeze-In becomes particularly tempting for models
        where the coupling strength required to generate correct relic
        density via freeze-out is ruled out by current exclusions from
        direct detection experiments. However, at this moment we shy
        away from the Freeze-In mechanism for the thermal relic. As the
        DM and the SM particles come from the same $E_6$ multiplets, it
        is not very straightforward to understand what causes the
        initial abundance of the DM to be negligible as compared to that
        of the SM. This $\sim 0$ initial abundance of the DM particles
        is a necessary initial condition for the freeze-in mechanism to
        work. Nevertheless, some studies of singlet Majorana DM with the
        freeze-in mechanism can be found in \cite{Klasen:2013ypa,
        Arcadi2018, Calibbi:2018fqf}.

         \begin{figure}[htpb]
           \centering
           \includegraphics{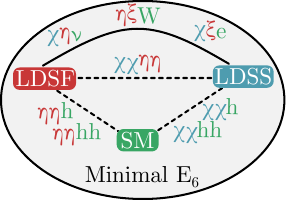}
           \caption{\textit{\small A schematic of the two-component DM paradigm as
           obtained from the fundamental of $E_6$, \textsf{E6DM}. The lightest particles of the scalar
           and the fermionic dark-sectors, the \LDSS\ and the \LDSF\ respectively, are
           simultaneously the DM candidates. The lighter of the two is absolutely
           stable, while the heavier is metastable and decays to the lighter one through
           unification scale and seesaw scale suppressed operators. The lifetime of the
           metastable DM is safe from current bounds. Both the DM particles annihilate
           to the SM through marginal and irrelevant couplings with the SM Higgs. The
           two sectors can annihilate into each other through a unification scale and
           seesaw scale suppressed operator, which we ignore.}}
         \label{fig:schema}
         \end{figure}

        The \LDSS\ behaves as a scalar singlet dark matter \cite{Guo:2010hq}. The
        effective Lagrangian (renormalisable), after EWSB, is given by
        \begin{align}
            \label{eq:SSRD}
            \ml\supset - m_\eta^2 \eta^2 - \mu_h \eta^2 h - \lambda \eta^2 h^2 - \lambda' \eta^4.
        \end{align}
        The trilinear and quartic interactions of $\eta$ with the SM
        Higgs, $h$, are generated from the quartics between  $h_u, h_d$
        and $\widetilde{\ell}$, as given in \cref{eq:scal_pot_1}.
        Therefore, $\eta$ either annihilates to two SM particles
        (including the Higgs) through the trilinear, or into two Higgses
        through the quartic. Detailed studies, including likelihood
        analyses, of such a DM particle can be found in
        \cite{GAMBIT:2017gge, Casas:2017jjg, GAMBIT:2018eea}. In \cite{Cline:2013gha},
        it is shown that the parameter space of a real scalar DM will be
        constrained by bounds from the XENON1T experiment,
        when the singlet contributes to all of the thermal relic.
        Indeed, in \cite{Arcadi:2019lka} the data from XENON1T have been
        taken into account  to show that the bound on the trilinear is
        as strong as $<0.001$ for 10 GeV $<M_\eta<$ 100 GeV.
        {\color{Black}Lastly, as is well known, if the scalar sector
            behaves like inert doublet dark matter, then a mass
            splitting among the CP-even and CP-odd particles $\sim 100$ keV is enough
        to suppress DM-nucleon couplings to an extent that makes the
        scenario safe
        from direct-detection constraints \cite{Hambye:2009pw}.}

        The above discussion on the nature of the stable and metastable
        DM candidates as obtained from the $E_6$ fundamental, along with
        the discussion in \cref{sec:symmbrk} on the symmetry strcture of
        the $E_6$ fundamental on restriction to $\gsm\times Z_2$ brings
        an end to our main points of interest. In \cref{fig:schema}, we
        describe the framework, labelled \textsf{E6DM},
        diagrammatically. In short, we describe a scenario of visible-
        and dark-sector unification, where a scalar DM sector, a fermion
        DM sector, and the SM (+ an RH neutrino) come from the same UV
        multiplet, three copies of which contain the three generations
        of the SM fermions and a single copy contains the EW breaking
        scalar. The lighter of the lightest dark-sector scalar (\LDSS)
        and the lightest dark-sector fermion (\LDSF) is absolutely
        stable, the heavier being metastable but with a lifetime allowed
        by all current bounds on decaying dark matter. The decay of the
        heavier DM particle to the lighter is unification scale, seesaw
        scale, and the electron Yukawa suppressed. The \LDSS\ and the
        \LDSF\ both annihilate to the SM through a Higgs portal. We have
        ignored the annihilation of the dark matter particles into each
        other. The most interesting point is that this framework only
        uses the multiplets required to get the SM from $E_6$ and do not
        introduce additional multiplets. In the next section,  we
        analyse the running of gauge couplings for a few representative
        symmetry breaking chains to show that the above paradigm can
        easily be embedded in minimal $E_6$ models for the most
        constrained case of one intermediate scale between
        $\cancel{M}_U$ and $\cancel{M}_\mathrm{EW}$ while satisfying the
        bounds set by proton-decay experiments and also while
        reproducing the correct light neutrino mass.

    \section{Symmetry breaking chains\label{sec:guts}}
        Having described all the moving parts that go into the \textsf{E6DM}
        paradigm, we look at the running of gauge couplings to establish that
        the hierarchy of masses that leads to the key results, as described
        above, can be embedded in a consistent model of gauge coupling
        unification (GCU). We stick to the minimal case of one scale,
        $\cancel{M}_I$, intermediate to the unification scale, $\cancel{M}_U$,
        and the electroweak scale, $\cancel{M}_\mathrm{EW}$.  This is the most
        stringent case, as adding more intermediate scales allows for more
        leeway to tune the parameters. To make our study representative, we look
        at one case each from the three routes given in \cref{fig:routes}, the
        $SU(3)\times SU(3)\times SU(3)$ route, the $SO(10)\times U(1)$ route,
        and the $SU(6)\times SU(2)$ route. As for scalar multiplets, we do not
        use multiplets with dimensionality larger than 351 in any way. As a
        result, we do not look at chains that require the extremely large 650 or
        1728 dimensional multiplets often used in $E_6$ model building. The
        three chains we look at are:
        \begin{subequations}
            \label{eq:chains}
        \begin{align}
            \mathrm{I}\quad&E_6\xrightarrow[M_U]{(15,1)_{62}\subset 351'}422
                \xrightarrow[M_I]{(10,1,3)_{422}\subset 351'}\gsm
                \xrightarrow[M_Z]{(1,2,\pm\half)_{\gsm}\subset 27}\gem,\\
            \mathrm{II}\quad&E_6\xrightarrow[M_U]{45(4)_{101}\subset 351}3221
                \xrightarrow[M_I]{(1,1,3)(-2)_{3221}\subset 351'}\gsm
                \xrightarrow[M_Z]{(1,2,\pm\half)_{\gsm}\subset 27}\gem,\\
            \mathrm{III}\quad&E_6\xrightarrow[M_U]{(1,1,8)_{333}\subset 78}3321
                \xrightarrow[M_I]{(1,6,3)(-\frac{2}{3})_{3321}\subset 351'}\gsm
                \xrightarrow[M_Z]{(1,2,\pm\half)_{\gsm}\subset 27}\gem.
        \end{align}
        \end{subequations}
        The first chain belongs to the $SU(6)\times SU(2)$  route of $E_6$
        breaking. This route is minimal in the sense that only the scalars
        required for giving masses to the SM fermions are used for the symmetry
        breaking of $E_6$ down to the SM. The sub-multiplet transforming as
        $(15,1)$ under $SU(6)\times SU(2)$, residing in the $351'$, breaks $E_6$
        down to the intermediate $SU(4)\times SU(2)\times SU(2)$ symmetry and
        simultaneously gives mass to $\widetilde{h}_{u/d}$ at the unification
        scale, resulting in the hierarchy of masses between the doublet
        Higgsinos and the singlet Higgsino, as described in \cref{sub:dsm}. The
        $\widetilde{h}_{u/d}$ reside in the $(\overline{6},2)$, with
        $(\overline{6},2)\times (\overline{6},2)\times (15,1)$ being a gauge
        singlet. For this chain, the unification scale is $1.3\times 10^{16}$
        GeV and the unified coupling strength is 0.55, resulting in a proton
        decay lifetime of $2.5\times 10^{36}$ years, that is allowed by current
        limits on the same from the Super-Kamiokande experiment, setting the
        exclusions lower limit at $\tau_p>1.6\times 10^{34}$ yrs
        \cite{Super-Kamiokande:2016exg}. The intermediate scale is at $1.4\times 10^{11}$
        GeV where $SU(4)\times SU(2)\times SU(2)$ breaks to $\gsm$, giving
        masses to the RH neutrinos. We discuss the details of the running of
        gauge couplings and the proton lifetime calculations below. Here, we
        note that both the unification scale and the intermediate is amenable to
        the paradigm of \textsf{E6DM}. Indeed, $\cancel{M}_I\sim 10^{11}$ GeV
        and $\cancel{M}_U\sim 10^{16}$ GeV are the `generic' values that we used
        in the last section to get our results. Rank breaking takes place in two
        steps. First, at the unification scale, $U(1)_D$ breaks to give mass to
        the doublet Higgsinos, and the second is at  the intermediate scale
        where the RH neutrinos get mass.

        \begin{table}[htpb]
            \centering
        \begin{tabular}{rllrrr}
            \toprule
            ID & Chain& $M_I$ (GeV)& $M_U$ (GeV)& $g_U$&$\tau_{p\to \pi^0 e^+}$ yrs\\
            \midrule
            I & {$4_C2_L2_R$} & $1.4\times 10^{11}$ & $1.3\times 10^{16}$ & 0.55& $2.5\times 10^{36}$\\
            II & {$3_C2_L2_R1_{BL}$} & $3.1\times 10^{9}$ & $1.2\times 10^{16}$ & 0.54 & $1.9\times 10^{36}$\\
            III & {$3_C3_L2_N1_N$} & $5.6\times 10^{6}$ & $4.9\times 10^{15}$ & 0.60& $3.6\times 10^{34}$\\
            \bottomrule
        \end{tabular}
        \caption{\textit{\small We write down the values of the intermediate and $E_6$ symmetry
        breaking scales for each of the chains in consideration. We  mention the value
        of the unified coupling constant, $g_U$. The last column shows the proton decay
        lifetime ($p\to \pi^0 e^+$) as calculated using $\cancel{M}_U$ and $g_U$.}
        \label{tab:uni}}
        \end{table}

        The second chain belongs to the $SO(10)\times U(1)$ route and is similar
        to the one discussed above. The unified symmetry is broken by the
        $(45)(4)_{101}$ sub-multiplet of $SO(10)\times U(1)$  that resides in
        the $351$ of $E_6$. The intermediate symmetry is broken to $\gsm$  by
        the $(1,1,3)(-2)_{3221}$ sub-multiplet residing in the $351'$ of $E_6$
        at a scale of $3.1\times 10^{9}$ GeV. This is the scale of RH neutrino
        mass. As can be checked from our order of magnitude arguments in
        \cref{sub:dna}, an intermediate scale around $10^9$ GeV, although lower
        than the generic $10^{11}$ GeV that we used, is still high enough for
        all the conclusions to remain valid. The doublet Higgsinos get masses
        from the vev of the $E_6$ breaking scalar (the mass term being the
        $(10,-2)^F_{101}\times (10,-2)^F_{101}\times (45,4)^S_{101}$ singlet).
        Therefore, they remain heavy at the unification scale, as required by
        \textsf{E6DM}. Although, the $U(1)_D$ breaking scalar, $s_4$,
        resides in the fundamental itself, we also get it from other multiplets
        that break the unification symmetry. For the first chain it was the
        $(15,1)_{62}$, whereas for this one, it is the $(45,4)_{101}\subset
        351$. This ensures that there isn't a huge hierarchy of vevs among the
        different directions of the fundamental. As a side note, the
        $SU(4)\times SU(2)\times SU(2)$ is a subgroup of both $SO(10)\times
        U(1)$ and $SU(6)\times SU(2)$. The reason we associate the $422$ chain
        with $SU(6)\times SU(2)$ and not $SO(10)\times U(1)$ is because of the
        $L\leftrightarrow R\, Z_2$ subgroup of $SO(10)$, commonly referred to as
        $D$-parity \cite{Chang:1983fu}. We checked that the one-step unification
        with $422D$ gives an $\cancel{M}_U$ that is too low to pass proton decay
        exclusions. $E_6$ is broken to $422$ through the $SO(10)$ route by the
        $(54)(4)_{101}\subset 351'$ which preserves $D$-parity. To break $E_6$
        to $422$ without D-parity, one needs to resort to the 650 dimensional
        representation that contains the 210-dim $SO(10)$ representation. As
        discussed above, we do not look at scalars in multiplets which are
        larger than 351 dimensional.

        The third chain is pathological in multiple ways and we will use it to
        underline a few key points. First of all, the intermediate scale is too
        low, $\sim 10^6$ GeV, for providing enough suppression to the operators
        that make the heavier DM particle to  decay to the lighter one. However,
        what is much more pressing is that both the additional ranks are broken
        at this intermediate scale. This implies that the doublet Higgsinos also
        get a mass at this scale. Since the first generation of the Higgsinos
        get a Yukawa suppression that is the same as electron Yukawas, there
        mass comes at around 1-10 GeV. {\color{Black}Moreover, the conventional type-I seesaw may not predict the tiny neutrino masses \cite{Esteban:2020cvm,DESI:2024mwx} together with the charged lepton masses, and quark masses and mixing \cite{Lazarides:1980nt} with such a low value of the intermediate scale.} Therefore, even from the point of view of
        single component dark matter, this chain is pathological. As discussed
        above, the Higgsinos can only get a mass at the scale where $U(1)_D$
        breaks. Therefore, for the \textsf{E6DM} paradigm to work, $U(1)_D$
        should break close to $\cancel{M}_U$, if not equal to it. It is
        important to note that not all SSBs break rank. The DM sector given by
        minimal $E_6$ is a result of the symmetry group itself. The
        two-component scenario arises when we choose a specific hierarchy
        between vevs. Check \cite{Schwichtenberg:2017xhv}, for a study of
        single-component fermionic DM arising from the $SO(10)$ chain of $E_6$.

        \begin{table}[t]
          \footnotesize
          \begin{subtable}{\linewidth}
          \centering
          \begin{tabular}{ccccc}
            \toprule
             \multicolumn{5}{c}{One- and two-loop $\beta$-coefficients for $\gsm\equiv SU(3)_c\times SU(2)_w\times U(1)_y$}\\
             $n_g=2$ &~ &$n_g=1$ & ~&$n_g=0$\\
             \midrule
             $\begin{pmatrix}-\frac{17}{3} \\ -\frac{11}{6} \\ \frac{163}{30}\end{pmatrix}
              \begin{pmatrix}-\frac{2}{3} & \frac{9}{2} & \frac{41}{30} \\
                12 & \frac{133}{6} & \frac{3}{2} \\
                \frac{164}{15} & \frac{9}{2} & \frac{667}{150}
              \end{pmatrix}$ &~&
              $\begin{pmatrix}-\frac{19}{3} \\ -\frac{5}{2} \\ \frac{143}{30}\end{pmatrix}
              \begin{pmatrix} -\frac{40}{3} & \frac{9}{2} & \frac{37}{30} \\
                12 & 14 & \frac{6}{5} \\
                \frac{148}{15} & \frac{18}{5} & \frac{316}{75} \end{pmatrix}$
                &~&
                $\begin{pmatrix}\frac{41}{10} \\-\frac{19}{6} \\-7 \end{pmatrix}
                \begin{pmatrix}
                   \frac{199}{50} & \frac{27}{10} & \frac{44}{5} \\
                   \frac{9}{10} & \frac{35}{6} & 12 \\
                   \frac{11}{10} & \frac{9}{2} & -26
                \end{pmatrix}$ \\
              \bottomrule
              \end{tabular}
              \caption{\textit{\small One- and two-loop $\beta$ coefficients for the
              three couplings of the $\gsm$ symmetry.}
              \label{tab:beta_coef_1}}
            \end{subtable}
            \vskip 1em
            \begin{subtable}{\linewidth}
              \centering
          \begin{tabular}{rlcc}
            \toprule
             Chain & Symmetry & \multicolumn{2}{c}{One- and two-loop $\beta$-coefficients}\\
             && $n_g=2$ & $n_g=1$\\
             \midrule
             I&$422$ & $\begin{pmatrix} -\frac{17}{3} \\ -1 \\ \frac{17}{3} \end{pmatrix},
                      \begin{pmatrix}
                        \frac{985}{6} & \frac{15}{2} & \frac{159}{2} \\
                        \frac{75}{2} & 33 & 6 \\
                        \frac{795}{2} & 6 & \frac{659}{3}
                      \end{pmatrix}$ &
                      $\begin{pmatrix}-\frac{19}{3} \\ -\frac{5}{3} \\ 5 \end{pmatrix},
                      \begin{pmatrix}
                        \frac{875}{6} & \frac{15}{2} & \frac{159}{2} \\
                        \frac{75}{2} & \frac{149}{6} & \frac{9}{2} \\
                        \frac{795}{2} & \frac{9}{2} & \frac{423}{2}
                      \end{pmatrix}$  \\
                      ~\\
              II&$3221$ &  $\begin{pmatrix} -\frac{17}{3} \\ -\frac{3}{2} \\ -\frac{5}{6} \\ \frac{22}{3} \end{pmatrix},
                          \begin{pmatrix} -\frac{2}{3} & \frac{9}{2} & \frac{9}{2} & \frac{7}{6} \\
                            12 & \frac{53}{2} & 6 & \frac{9}{4} \\
                            12 & 6 & \frac{271}{6} & \frac{57}{4} \\
                            \frac{28}{3} & \frac{27}{4} & \frac{171}{4} & \frac{401}{12}
                          \end{pmatrix}$ &
                        $\begin{pmatrix} -\frac{19}{3} \\ -\frac{13}{6} \\ -\frac{3}{2} \\ \frac{20}{3} \end{pmatrix},
                        \begin{pmatrix} -\frac{40}{3} & \frac{9}{2} & \frac{9}{2} & \frac{5}{6} \\
                          12 & \frac{55}{3} & \frac{9}{2} & \frac{9}{4} \\
                          12 & \frac{9}{2} & 37 & \frac{57}{4} \\
                          \frac{20}{3} & \frac{27}{4} & \frac{171}{4} & \frac{397}{12}
                        \end{pmatrix}$ \\
                        ~\\
              III&$3321$ &  $\begin{pmatrix}-5 \\ -\frac{7}{6} \\ \frac{19}{6} \\ \frac{67}{6} \end{pmatrix},
                        \begin{pmatrix} 12 & 12 & \frac{9}{2} & \frac{3}{2} \\
                                        12 & \frac{529}{3} & \frac{135}{2} & \frac{155}{6} \\
                                        12 & 180 & \frac{440}{3} & 18 \\
                                        12 & \frac{620}{3} & 54 & \frac{170}{3}
                        \end{pmatrix}$ & -- N/A -- \\
              \bottomrule
              \end{tabular}
          \caption{\textit{\small One- and two-loop $\beta$ coefficients for the
              different intermediate symmetries for different cardinalities of the Higgsinos and the heavy quarks.}\label{tab:beta_coef_2}}
        \end{subtable}
        \caption{\textit{\small One- and two-loop $\beta$-coefficients
        corresponding to the different gauge couplings. The two-loop
        coefficients are calculated using the techniques in \cite{Jones:1981we,
        Machacek:1983tz}. The column matrices give the one-loop coefficients and
        the $2\times 2$ matrices give the two-loop coefficients. The ordering of
        the coefficients follow the ordering of the symmetries in their naming.
        For example, for $\gsm$, $SU(3)_c$ is followed by $SU(2)_w$ followed by
        $U(1)_y$. The quantity $n_g$ gives the number of generations of the
        Higgsinos and the Heavy Quarks that contribute to the running at that
        stage.} \label{tab:beta_coef}}
        \end{table}

        To get the scales and the couplings, as shown in \cref{tab:uni},
        we calculate the running of the gauge couplings using two-loop
        $\beta$-coefficients, following the method outlined in
        \cite{Bandyopadhyay:2015fka, Chakrabortty:2017mgi,
        Chakrabortty:2019fov}. The $\beta$-coefficients corresponding to
        the different symmetry groups and the different stages of
        running are given in \cref{tab:beta_coef}. Note
            that, we do not consider the contribution to the
        \(\beta\)-coefficients from Yukawa interactions as the
    contribution from these are typically less that one
percent~\cite{Bertolini:2009qj}. For the first two chains, all three
generations of the doublet Higgsinos and the heavy quarks get mass from
the $E_6$ breaking vev. The masses  of the three different generations
are Yukawa suppressed, the Yukawa couplings being commensurate with the
Yukawa couplings of the three generations of the SM fermions, as
discussed in \cref{sub:dsm}. Accordingly, we keep the third generation
at the unification scale itself, the second generation is integrated out
at $10^{-3}\times \cancel{M}_U$ and the first generation at
$10^{-6}\times \cancel{M}_U$. There will be slight differences in the
Yukawas due to running (of these Yukawa couplings), but we have checked
that small changes to the masses of these fermions do not modify the
resulting scales in any significant way. As the mass of the first
generation (lightest) is close (within an order of magnitude) to the
intermediate scale for the first two chains, we do not introduce its
mass as a different scale, but as threshold corrections
\cite{Weinberg:1980wa,Hall:1980kf} corresponding to the intermediate
scale. Therefore, between the unification and the intermediate scale,
there is another scale at $10^{-3}M_U$, obtained by integrating out the
second generation of the doublet Higgsinos and the heavy quarks. The
third generation is kept at the unification scale and the first
generation is introduced as threshold correction to the intermediate
scale.  In \cref{tab:beta_coef_2}, we separately show the
$\beta$-coefficients corresponding to the intermediate symmetries where
the cardinality ($n_g$) of the dark-fermion generations is two and where
it is one. The matching at the intermediate scale for the couplings are
given by:
        \begin{subequations}
        \begin{alignat}{2}
            \mathrm{(I)}\;&  &&w_c = w_4 - \frac{1}{3\pi} - \mathcal{R}_\mathrm{T};\;
            w_w = w_2 -\mathcal{R}_\mathrm{T};\;
            w_y = \frac{3}{5} w_{2N} + \frac{2}{5} w_4 -\frac{7}{30\pi}-\mathcal{R}_\mathrm{T},\\
            \text{(II)}\;& &&w_c= w_3 -\mathcal{R}_\mathrm{T};\;
            w_w = w_2 -\mathcal{R}_\mathrm{T};\;
            w_y = \frac{3}{5} w_R + \frac{2}{5}w_{BL} - \frac{1}{6\pi} - \mathcal{R}_\mathrm{T},\\
            \text{(III)}\; & &&w_c = w_c;\;
             w_w = w_{3W} - \frac{1}{12\pi};\;
             w_y = \frac{3}{5} w_{2N} + \frac{1}{5} w_N  + \frac{1}{5} w_{3W} -\frac{3}{20\pi},\\
              \mathrm{with},\,& &&w_i= \frac{g_i^2}{4\pi};\; \mathcal{R}_\mathrm{T}= \frac{1}{3\pi}\left(\ln\frac{M_U}{M_I}-6\ln 10\right).
        \end{alignat}
        \end{subequations}
        The quantity $\mathcal{R}_\mathrm{T}$ is the threshold correction as
        discussed above. For the first two chains, only the singlet dark matter
        particles and the singlet scalar $\phi$ (the mass eigenstate of $s_8$)
        remain propagating degrees of freedom on top of the SM particles (as the
        sleptons are kept heavy at the intermediate scale, \cref{sub:dsm}) below
        the intermediate scale. As gauge singlets do not contribute to
        $\beta$-coefficients of the gauge couplings, the $\gsm$ running of
        couplings for the first two chains are  the same as the SM case, as can
        be seen from column 3 of \cref{tab:beta_coef_1}. For the disallowed
        chain III, all three generations contribute to running between
        $\mathcal{M}_U$ and $\mathcal{M_I}$. The Higgsinos and the heavy quarks
        are integrated out in the SM stage of running and hence we show the
        $\beta$ coefficients of the SM symmetry for the two stages corresponding
        to 2 generations and 1 generation of the exotic fermions propagating
        ($n_g=2$ and $1$ respectively) in \cref{tab:beta_coef_1}. The hierarchy
        of the masses for the first two chains (i.e. the allowed chains) is
        schematically shown in \cref{fig:Scal}.

        In non-supersymmetric models of unification, gauge boson induced dim-6 operators
        are the leading contributors to proton decay. These operators are:
        \cite{FileviezPerez:2004hn, Nath:2006ut, Babu:2015bna,}:
        \begin{subequations}
        \begin{align}
            \label{operator_physical_basis}
            \mathcal{O}_L\left( e^c, d\right) & = \frac{g_U^2}{2 M_U^2} \left[ 1 +
                |V_{ud}|^2\right]\;  \epsilon^{ijk} \overline{u^c_i}\gamma^\mu u_j \overline{e^c} \gamma_\mu d_{k} , \\
            \mathcal{O}_R\left( e, d^c\right) & = \frac{g_U^2}{2 M_U^2} \left[ 1 +
                |V_{ud}|^2\right]\; \epsilon^{ijk}
            \overline{u^c_i}\gamma^\mu u_j \overline{d^c_{k}} \gamma_\mu e,
        \end{align}
        \end{subequations}
        where $|V_{ud}| = 0.9742$ is the CKM matrix element \cite{Tanabashi:2018oca}
        between up and down quarks. The partial width for the channel $p\to \pi^0 e^+$
        is expressed as \cite{Babu:2015bna,FileviezPerez:2004hn}:
        \begin{align}
            \frac{1}{\tau_p}& =\frac{m_p}{32\pi}\left(1-\frac{m_{\pi^0}^2}{m_p^2}\right)^2 R_L^2 \frac{g_U^4}{4 \cancel{M}_U^4}(1+|V_{ud}|^2)^2
         \left( R_{SR}^2 |\langle \pi^0 \rvert (ud)_R u_L\lvert p \rangle |^2 +
            R_{SL}^2 |\langle \pi^0 \rvert (ud)_L u_L\lvert p \rangle |^2
         \right),
        \end{align}
        where $m_p$=938.3 MeV and $m_{\pi^0}$= 134.98 MeV are the masses of
        proton and the neutral pion respectively \cite{Tanabashi:2018oca}.
        $R_L\simeq 1.33$ \cite{Nihei:1994tx} is the long-range renormalisation
        factor for the proton decay operator from the electroweak scale to the
        QCD scale ($\sim 1$ GeV), whereas, $R_{SR(L)}$ is the short-range
        enhancement factor arising due the renormalisation group evolution of
        the proton decay operator $\mathcal{O}_{R(L)}$ from the unification to
        the electroweak scale. The short-range enhancement factors depend on the
        breaking chains. However, we use a conservative value,
        $R_{SR}=R_{SL}=2.4$  \cite{Buras:1977yy,Abbott:1980zj}. The form-factors
        as provided by the lattice QCD computation are \cite{Aoki:2017puj}:
        \begin{align}
        \langle \pi^0 \rvert (ud)_R u_L\lvert p \rangle  = -0.131, \ \ \langle \pi^0 \rvert (ud)_L u_L\lvert p \rangle = 0.134 \ .
        \end{align}
        As a final check of the two `allowed' chains, we
            compute the coupling at \(M_\mathrm{Pl}\). For the
            \(4_C2_L2_R\) route we find \(g_U(M_\mathrm{Pl})\ \simeq\
            0.57\) and for the \(3_C2_L2_R1_{BL}\) chain we find
            \(g_U(M_\mathrm{Pl})\ \simeq\ 0.86\).  Therefore, in both
            the cases the coupling is perturbative till the Planck
            scale. For the \(4_C2_L2_R\) chain, the one and two loop
        \(\beta\) coefficients are -9 and 5956 respectively. For the
    \(3_C2_L2_R1_{BL}\) chain, -5 and 6628 are the one- and the two-loop
\(\beta\) coefficients. It is worth mentioning that we assume that the symmetry breaking patterns in \ref{eq:chains}, which algebraically seems feasible \cite{Slansky:1981yr}, can be tuned from the general potential of the two or more multiplets. Otherwise, we need to introduce higher multiplets such as 650, which can spoil the perturbativity of $g_U$ till the Planck scale. However, the running of gauge couplings below the GCU scale and hence the phenomenology remains the same.

        In summary, we have shown that the hierarchy of masses and scales as
        required by the \textsf{E6DM} paradigm can be achieved easily in models
        of unification, using a few representative chains of GCU. We have also
        used a chain where the paradigm does not work to establish the relation
        between the masses of the exotic fermions and rank breaking of $E_6$. We
        have calculated the proton decay lifetime for the chains under
        consideration to show that they are allowed by current exclusions. The
        hierarchy of masses of the scalars are not protected against radiative
        corrections, like all non-supersymmetric models with multiple stages of
        symmetry breaking. However, the models can be supersymmetrised to
        stabilise the masses. Our results are all based on the gauge symmetry
        under question and will not interfere with global supersymmetry. We have
        intentionally restricted ourselves to the minimal case of one
        intermediate symmetry and to small scalar multiplets as these are the
        cases with least freedom of playing around with parameters. Therefore,
        the results shown here should easily be reproduced in non-minimal models
        with more symmetry breaking scales and larger scalar multiplets.

    \section{Conclusion\label{sec:conclusion}}%
    We discuss a framework for dark- and visible-sector unification with
    the SM embedded in $E_6$, extending the work presented in
    Ref.~\cite{Schwichtenberg:2017xhv}. The copies of the $E_6$
    fundamental containing the SM include the dark-sector as well. The
    framework accommodates multicomponent dark matter, one scalar and
    one fermionic. The lighter of the two is stable, while the heavier
    is metastable with a lifetime allowed by current exclusions. The
    stability of the heavier component is ensured by the $E_6$ symmetry
    and the hierarchy of vevs. The framework is not partial to any
    specific symmetry breaking route of $E_6$. As we discussed a
    non-SUSY version of $E_6$, the hierarchical separation between the
    scalars is not guaranteed to be preserved by radiative corrections.
    The simplest solution to this is to supersymmetrise the model
    \cite{King:2005my, Babu:2015psa}. Our results follow from the gauge
    symmetry and will not be modified for a SUSY version. The
    calculations for the running of couplings, however, would be
    modified.

    We discuss, in detail, the different ways in which the SM
    hypercharge is obtained from a linear combination of the three
    additional ranks of $E_6$ (ones in addition to the non-abelian part
    of the SM). We show that from this definition of hypercharge we can
    separate the chains that can give rise to a DM stabilising $Z_2$
    symmetry at the IR from the ones that can't. For the definitions
    that preserve this discrete symmetry, the particle content
    (SM+exotics) transforms identically under the $SU(3)_c\times
    SU(2)_w\times U(1)_y\times U(1)_D\times U(1)'$, irrespective of
    intermediate symmetry breaking. Here, $U(1)_D$ and $U(1)'$ are the
    $U(1)$ symmetries corresponding to the additional ranks of $E_6$. It
    is $U(1)_D$ that differentiates the SM particles from the
    dark-sector particles, and its discrete remnant stabilises the DM
    particles at the IR.

    The scalar DM resides in the multiplet that contains the EW breaking
    Higgs doublets, and the fermionic DM comes from the multiplets
    containing the SM fermions. The fermions being embedded in the
    anomaly free fundamental of $E_6$, the question of anomaly
    cancellation does not arise. What is notable is that there is
    intra-SM anomaly cancellation for the $U(1)'$ that doesn't stabilise
    the DM and there is SM-DM mixed anomaly cancellation for the DM
    stabilising $U(1)_D$. 

    The interactions between the lightest particles of the fermionic and
    the scalar sectors are all suppressed by the unification scale and
    the electron Yukawa simultaneously. The resulting decay width is
    large enough to give a lifetime that is much larger than the age of
    the Universe. The hierarchy that guarantees the metastability of the
    heavier DM particle is a choice. It is possible to build models with
    a different choice of scales where only the lightest particle is a
    DM candidate, dictated by the stabilising remnant of $U(1)_D$.

    At the IR, both the scalar and the fermionic DM particles annihilate
    to the SM through the Higgs portal. The fermionic DM talking to the
    Higgs through the Yukawa coupling that gives it mass. Our interest
    was in the qualitative features of this two-component DM sector. We
    have discussed the thresholds that go into the spectrum, mentioning
    the different symmetry breaking scales which determine the masses of
    the exotics. We can extend this general framework to build various
    models with diverse IR phenomenology.  We look at the two stable DM
    candidates individually, but a proper study of the cosmological
    evolution, including the interplay of the two different DM
    candidates, is an interesting prospect. However, as the couplings of
    the DM particles to each other are all high-scale suppressed, it is
    a good approximation to treat them separately. Also, any proper
    phenomenological study based on a Unified framework must include a
    multi-parameter fit involving observed values of the fermion masses,
    low energy observables, and running of the scalar and Yukawa
    couplings leading to unification. That was certainly not our goal,
    and we leave this exercise for future endeavours. Here, as proof of
    principle, we study the unification of gauge couplings for the three
    maximal subgroups of $E_6$. As seen from this exercise, we can
    easily embed the framework in a minimal model of gauge coupling
    unification. We also describe a `null' case where the setup breaks
    down.

    \textbf{Acknowledgements: }\textit{The authors thank Avik Banerjee
        and Joydeep Chakrabortty for useful discussions and also for
        giving the final manuscript a thorough read. TB acknowledges the
        Workshop on High Energy Physics Phenomenology (WHEPP) 2017 for
        providing a platform for discussions that eventually led to this
        work. TB also thanks Joydeep Chakrabortty for hosting him at IIT
        Kanpur, India for two weeks in March 2018 where this work
        commenced. RM is supported by IBS under the project code:
        IBS-R018-D3. }

    \begin{appendix}
        \section{\texorpdfstring{\pd\ for all the E\textsubscript{6}
        chains}{PD for all the E6 chains} \label{sec:E6}}%
        In this appendix, we explicitly show the decomposition of the $E_6$
        fundamental under the different intermediate symmetries. As discussed in
        the main text, there are three distinct maximal subgroups of $E_6$
        ---$SU(3)\times SU(3)_L\times SU(3)_N$, $SO(10)\times U(1)$, and
        $SU(6)\times U(1)$ (see \cref{fig:routes})--- that can reproduce the
        Standard Model (SM). Each of these subgroups has multiple chains of
        symmetry breaking taking it to $\gsm$. In \cref{tab:tab1}, we
        demonstrate how the fundamental transforms under the intermediate
        $SU(3)\times SU(2)\times U(1)_{x_1}\times U(1)_{x_2}\times U(1)_{x_3}$
        when obtained from the different subgroups. The table also indicates the
        linear combinations, $(C_1, C_2, C_3)$, of $(x_1,x_2,x_3)$ that
        reproduce the SM hypercharges. We do not show the orthogonal
        combinations explicitly. Note, we can perform any orthogonal
        transformation in the $y',y''$ plane, keeping the physics at the IR the
        same. We have not shown `GUT-normalised' \cite{Georgi:1974yf}
        values of the charges anywhere in the text. However, while calculating
        the $\beta$-coefficients we have used the GUT-normalised values.

        \begin{table}[htpb]
            \small
            \centering
            \begin{tabularx}{\textwidth}{lll}
                \toprule
                \toprule
                Num&Symmetry & Decomposition of \underline{27} of $E_6$\\
                \midrule
                (1)&$3_C3_W3_N$&(3,3,1)+($\overline{3},1,\overline{3}$)+(1,$\overline{3}$,3)\\
                &$3_C2_W3_N(1_W)$ & [(3,2,1)(1)+(3,1,1)(-2)]
                +[($\overline{3},1,\overline{3}$)(0)] + [(1,$2^*$,3)(-1) + (1,1,3)(2)]
                \tabularnewline
                &$3_C2_W2_N(1_W1_N)$&[(3,2,1)(1,0)+(3,1,1)(-2,0)]
                    +[\{($\overline{3}$,1,$2^*$)(0,-1)+($\overline{3}$,1,1)(0,2)\}]  \\
                &&+[\{(1,$2^*$,2)(-1,1)+(1,$2^*$,1)(-1,-2)\}+\{(1,1,2)(2,1) +(1,1,1)(2,-2)\}]\\
                &$3_C2_W(1_W1_N1_R)$&[(3,2)(1,0,0)+(3,1)(-2,0,0)]
                    +[\{($\overline{3}$,1)(0,-1,$\pm$1/2) +
                    ($\overline{3}$,1)(0,2,0)]  \\
                && + [\{(1,$2^*$)(-1,1,$\pm$ 1/2)+(1,$2^*$)(-1,-2,0)\}
                    +\{(1,1)(2,1,$\pm$ 1/2)]+(1,1)(2,-2,0)\}] \\~\\
                &($C_1,C_2,C_3$):& $ (1/6,1/6,\pm 1),(1/6,-1/3,0)$\\
                \midrule[0.1em]
                (2)&$SO(10)1_6$&16(1)+10(-2)+1(4)\\
                &5$(1_61_{10})$&$[\overline{5}$(1,3)+10(1,-1)+1(1,-5)]
                    + [5(-2,2)+$\overline{5}$(-2,-2)] + [1(4,0)]\\
                &$32 (1_6 1_{10}1_5)$&
                [(1,$2^*$)(1,3,-3) +
                ($\overline{3}$,1)(1,3,2) + (3,2)(1,-1,1) +
                ($\overline{3}$,1)(1,-1,-4) + (1,1)(1,-1,6)]\\
                &&+(1,1)(1,-5,0)]+[(1,2)(-2,2,3) + (3,1)(-2,2,-2)
                  +(1,$2^*$)(-2,-2,-3) \\
                &&+ ($\overline{3}$,1)(-2,-2,2)]+[(1,1)(4,0,0)]\\~\\
                &$(C_1, C_2,C_3)$:& $
                (0,0,1/6),(0,-1/5,-1/30),(1/4,1/20,-1/30)$\\
                \midrule
                (3)&$422(1_6)$&[($\overline{4}$,2,1)(1)]+[(4,1,2)(1)]
                    +[(1,2,2)(-2)]+[(6,1,1)(-2)]+[(1,1,1)(4)]\\
                &$322(1_61_{B-L})$&
                [($\overline{3}$,2,1)(1,1/3) + (1,2,1)(1,-1)] +
                [(3,1,2)(1,-l/3) +(1,1,2)(1,1)]\\
                && + [(1,2,2)(-2,0)]+[(3,1,1)(-2,2/3)+($\overline{3}$,1,1)(-2,-2/3)]
                + [(1,1,1)(4,0)]\\
                &$32(1_61_{B-L}1_R)$&
                [($\overline{3}$,1)(1,1/3,$\pm$1/2)
                    +(1,1)(1,-1,$\pm$1/2)]
                    + [(3,2)(1,$-1/3$,0) +(1,2)(1,1,0)]\\
                &&  + [(1,2)(-2,0,$\pm$1/2)]+[(3,1)(-2,2/3,0)+($\overline{3}$,1)(-2,-2/3,0)]
                    +[(1,1)(4,0,0)]\\
                    \\
                &$(C_1, C_2, C_3)$:& $ (0,-1/2,\pm 1),(1/4,1/4,0)$\\
                \midrule[0.1em]
                (4)&6$2$&$(\overline{6},2)+(15,1)$\\
                &5$2$($1_\chi$) &
                $[(1,2)(5)+(\overline{5},2)(-1)]+[(5,1)(-4)+(10,1)(2)]$\\
                \cmidrule(l{10px}r{30px}){1-3}
                (4a) &3$2_52$($1_\chi  1_\xi$)&
                $[(1,1,2)(5,0)
                 + \{(1,2^*,2)(-1,-3) + (\overline{3},1,2)(-1,2)\}]
                 + [\{(1,2,1)(-4,3)$\\
                &&$+(3,1,1)(-4,-2)\}+\{(1,1,1)(2,6)+(3,2,1)(2,1)+(\overline{3},1,1)(2,-4)\}]$\\
                &3$2_5$($1_\chi 1_\xi 1_\zeta$) &
                $[(1,1)(5,0,\pm \frac{1}{2})+ \{(1,2^*)(-1,-3,\pm
                    \frac{1}{2})+(\overline{3},1)(-1,2,\pm
                    \frac{1}{2})\}]$\\
                &&$+[\{(1,2)(-4,3,0)+(3,1)(-4,-2,0)\}+\{(1,1)(2,6,0)+(3,2)(2,1,0)$\\
                &&$+(\overline{3},1)(2,-4,0)\}]$\\~\\
                &$(C_1, C_2,C_3)$:& $(0,1/6,0), ({1}/{10},-{1}/{30},\pm 1)$\\
                \cmidrule(l{10px}r{30px}){1-3}
                (4b)&32($1_\chi 1_\xi 1_\lambda$) &
                $[(1,2)(5,0,0)+
                \{(1,2)(-1,-3,\mp\half)+(\overline{3},2)(-1,2,0)\}]+[\{(1,1)(-4,3\pm\half)$\\
                    &&$+(3,1)(-4,-2,0)\}+\{(1,1)(2,6,0)+(3,1)(2,1,\pm\half)
                +(\overline{3},1)(2,-4,0)\}]$\\
                 ~\\
                &$(C_1, C_2,C_3)$:& $ ({1}/{10},-{1}/{30}, \pm 1),(-1/10,-2/15,0)$\\
                \cmidrule(l{10px}r{30px}){1-3}
                (4c) &42($1_5 1_\chi$) &
                $[(1,2)(0,5)+(\overline{4},2)(1,-1)+(1,2)(-4,-1)]+[(1,1)(4,-4)+(4,1)(-1,-4)]$\\
                && $+[(4,1)(3,2)+(6,1),(-2,2)]$\\
                &32($1_4 1_5 1_\chi$) &
                $[(1,2)(0,0,5)+(\overline{3},2)(1/3,1,-1)+(1,2)(-1,1,-1)+(1,2)(0,-4,-1)]$
                \\&&
                $+(1,1)(1,-1,-4)+(3,1)(-1/3,-1-4)]+[(3,1)(-1/3,3,2)+(1,1)(1,3,2)$\\
                &&$+(3,1)(2/3,-2,2)
                +(\overline{3},1)(-2/3,-2,2)]+[(1,1)(0,4,-4)$\\
                ~\\
                &$(C_1,C_2,C_3)$ & $(-1/2,\pm 1/10,\pm 1/10),(1/4,-3/20,1/10)$
                 \\
                \midrule
                (5)&6$2$ & $(\overline{6},2)+(15,1)$\\
                &422(1) &
                $[(1,2^*,2)(-2)+(\overline{4},1,2)(1)]+[(1,1,1)(4)+(6,1,1)(-2)+(4,2,1)(1)]$ [$\to$ (3)]\\
                \midrule
                (6)&6$2$ & $(\overline{6},2)+(15,1)$\\
                &332(1) &$ [(\overline{3},1,2)(-1)+(1,\overline{3},2)(1)] +
                [(3,3,1)(0)+(\overline{3},1,1)(2)+(1,\overline{3},1)(-2)]$ [$\to$ (1)]\\
                \bottomrule
                \bottomrule
            \end{tabularx}
            \caption{\small The $E_6$ fundamental on restriction to the different subgroups.
            We have appropriately used parentheses, braces, and brackets to facilitate
            the reading of the appropriate decompositions. We have provided the
            appropriate linear combinations of the three abelian charges that give
            the hypercharge for each chain. See text for details.
            \label{tab:tab1}}
        \end{table}

        The second column in \cref{tab:tab1} indicates the symmetry breaking
        chain, the first, second, and third rows representing the $SU(3)^3$
        route, the $SU(5)$ chain of the $SO(10)$ route, and the Pati-Salam
        \cite{Pati:1973rp, Pati:1974yy} chain of the $SO(10)$ route respectively. Note that the
        $SU(5)$ route of $SO(10)$ is the only case where two of the three
        possible hyper-charge definitions are not `mirrors' of each other. The
        rest of the table deals with the $SU(6)\times SU(2)$ route of descent.
        In row 4, $SU(6)$ is broken into $SU(5)\times U(1)$, accounting for one
        of the three $U(1)$ symmetries. The other two $U(1)$ symmetries can come
        in three different ways, as shown in sub-rows 4a, 4b, and 4c. In
        sub-rows 4a and 4b, the $SU(5)$ is broken to $SU(3)\times SU(2)\times
        U(1)$. In 4a, the $SU(2)$ in $SU(5)$ remains intact while the $SU(2)$
        present from the beginning (the $SU(2)$ of $SU(6)\times SU(2)$) breaks
        to give the third $U(1)$. In 4b, the $SU(2)$ coming from $SU(5)$ breaks
        to give the third $U(1)$. In 4c, $SU(5)$ breaks to $SU(4)\times U(1)$
        and $SU(4)$ breaks to $SU(3)\times U(1)$, giving the second and third
        $U(1)$ symmetries. The fifth row shows the case where $SU(6)$ breaks to
        $SU(4)\times SU(2)$. The charges, in this case, map to the Pati-Salam
        case of $SO(10)$ (row 3). Similarly, the sixth row where $SU(6)$ breaks
        to $SU(3)\times SU(3)$ maps to the trinification chain (row 1).

        \begin{table}[htpb]
            \footnotesize
          \centering
        \begin{tabular}{rlll}
        \toprule
                SI& \pd &$(C_1,C_2,C_3)$& Decomposition of \underline{27} of $E_6$\\
        \midrule
                  (1.i)&$\checkmark$&$\left (\frac{1}{6},\frac{1}{6},\pm 1\right)$&
                    $
                      \left[(3,2,\frac{1}{6})(1,1)
                      + (\overline{3},1,-\frac{2}{3})(1,-\frac{5}{6})
                      + (\overline{3},1,\frac{1}{3})(1,-\frac{7}{6})
                      + (1,2^*,-\frac{1}{2})(1,-3)\right.
                    $
                    \\ &&&
                    $\left.
                      + (1,1,1(1,\frac{17}{6})
                      + (1,1,0)\left(1,\frac{19}{6}\right)\right]
                      + \left\{
                      (1,1,0)(4,0)
                      + (1,2^*,\pm\frac{1}{2})(-2,\mp\frac{1}{6})
                      \right\}
                    $
                    \\&&&
                    $
                      + (\overline{3},1,\frac{1}{3})(-2,2)
                      + (3,1,-\frac{1}{3})(-2,-2)
                    $
                   \\~ \\
                  (1.ii)& $\times$&$\left(\frac{1}{6},-\frac{1}{3},0\right)$ &
                    $
                      (3,2,\frac{1}{6})(0,1)
                      + (\overline{3},1,-\frac{2}{3})(0,1)
                      + (1,1,1)(0,1)
                      + (1,2^*,\half)(0,-2)
                     $
                  \\&&&
                  $
                    + (1,2^*,-\half)(\half,-\half)
                    + (1,2^*,-\half)(-\half,-\half)
                    + (\overline{3},1,\frac{1}{3})(\half,-\half)
                  $
                  \\&&&
                  $
                    + (\overline{3},1,\frac{1}{3})(-\half,-\half)
                    + (1,1,0)(\half,\frac{5}{2})
                    + (1,1,0)(-\half,\frac{5}{2})
                    + (3,1,-\frac{1}{3})(0,-2)
                  $
                   \\
        \midrule
                (2.i)& $\checkmark$&$(0,0,\frac{1}{6})$&
                $
                \left[
                    (3,2,\frac{1}{6})(1,1)
                  + (\overline{3},1,-\frac{2}{3})(1,1)
                  + (\overline{3},1,\frac{1}{3})(1,-3)
                  + (1,2,-\half)(1,-3)
                \right.
                $
                \\&&&
                $
                \left.
                  + (1,1,1)(1,1)
                  + (1,1,0)(1,5)
                \right] + \left\{
                  (1,2,\pm\half)(-2,\mp 2)
                  + (1,1,0)(4,0)
                \right\}
                $
                \\&&&
                $
                  + (\overline{3},1,\frac{1}{3})(-2,2)
                  + (3,1,-\frac{1}{3})(-2,-2)
                $
                \\~\\
                (2.ii)&$\checkmark$ &$\left(0,-\frac{1}{5},-\frac{1}{30}\right)$&
                $ \left[
                (3,2,\frac{1}{6})(1,-\frac{7}{6})
                + (\overline{3},1,-\frac{2}{3})(1,-\frac{3}{2})
                + (\overline{3},1,\frac{1}{3})(1,\frac{23}{6})
                 + (1,2,-\half)(1,\frac{7}{2})
                \right.$
                \\&&&
                $\left.
                 + (1,1,1)(1,-\frac{5}{6})
                 + (1,1,0)(1,-\frac{37}{6})
                \right] + \left\{
                 (1,2,\pm\half)(-2,\pm\frac{8}{3})
                 + (1,1,0)(4,0)
                \right\}$
                \\&&&
                $
                 + (\overline{3},1,\frac{1}{3})(-2,-\frac{7}{3})
                 + (3,1,-\frac{1}{3})(-2,\frac{7}{3})
                $
                \\~\\
                (2.iii)&$\times$ &$\left(\frac{1}{4},\frac{1}{20},-\frac{1}{30}\right)$&
                $ (3,2,\frac{1}{6})(4,\frac{9}{4})
                 + (\overline{3},1,-\frac{2}{3})(6,\frac{29}{10})
                 + (1,1,1)(2,\frac{8}{5})
                 + (1,2,\half)(-10,-\frac{103}{20})
                $
                \\&&&
                $
                 + (1,2,-\half)(7,-\frac{27}{4})
                 + (1,2,-\half)(-9,\frac{103}{20})
                 + (\overline{3},1,\frac{1}{3})(-11,\frac{9}{2})
                $
                \\&&&
                $
                 + (\overline{3},1,\frac{1}{3})(5,-\frac{37}{5})
                + (1,1,0)(19,0)
                + (1,1,0)(3,\frac{119}{10})
                + (3,1,-\frac{1}{3})(-8,-\frac{9}{2})
                $
                \\
        \midrule
                (3.i)&$\checkmark$ &$\left(0,-\frac{1}{2},\pm 1\right)$&
                $\left[
                   (3,2,\frac{1}{6})(1,-\frac{1}{3})
                  + (\overline{3},1,-\frac{2}{3})(1,\frac{1}{12})
                  + (\overline{3},1,\frac{1}{3})(1,\frac{7}{12})
                  + (1,2,-\half)(1,1)
                \right.$
                \\&&&
                $\left.
                  + (1,1,1)(1,-\frac{3}{4})
                  + (1,1,0)(1,-\frac{5}{4})
                \right] + \left\{
                  (1,2,\pm\half)(-2,\pm\frac{1}{4})
                  + (1,1,0)(4,0)
                \right\}
                $
                \\&&&
                 $
                  + (\overline{3},1,\frac{1}{3})(-2,-\frac{2}{3})
                  + (3,1,-\frac{1}{3})(-2,\frac{2}{3})
                 $\\~\\
                (3.ii)&$\times$ &$\left(\frac{1}{4},\frac{1}{4},0\right)$ &
                  $
                    (3,2,\frac{1}{6})(0,\frac{4}{3})
                    + (\overline{3},1,-\frac{2}{3})(0,-\frac{4}{3})
                    + (1,1,1)(0,4)
                    + (1,2,\half)(0,0)
                  $
                \\&&&
                  $
                    + (1,2,-\half)(\half,-2)
                    + (1,2,-\half)(-\half,-2)
                    + (\overline{3},1,\frac{1}{3})(\half,\frac{2}{3})
                  $
                \\&&&
                  $
                    + (\overline{3},1,\frac{1}{3})(-\half,\frac{2}{3})
                    + (1,1,0)(\half,2)
                    + (1,1,0)(-\half,2)
                    + (3,1,-\frac{1}{3})(0,-\frac{8}{3})
                  $
                \\
        \midrule
                (4a.i)&$\times$ &$\left(0,\frac{1}{6},0\right)$ &
                $
                  (3,2,\frac{1}{6})(2,0)
                  + (\overline{3},1,-\frac{2}{3})(2,0)
                  + (1,1,1)(2,0)
                  + (1,2,\half)(-4,0)
                $
                \\&&&
                $
                  + (1,2,-\half)(-1,\half)
                  + (1,2,-\half)(-1,-\half)
                  + (\overline{3},1,\frac{1}{3})(-1,\half)
                $
                \\&&&
                $
                  + (\overline{3},1,\frac{1}{3})(-1,-\half)
                  + (1,1,0)(5,\half)
                  + (1,1,0)(5,- \half)
                  + (3,1,-\frac{1}{3})(-4,0)
                $
                \\~\\
                (4a.ii)&$\checkmark$&$\left(\frac{1}{10},-\frac{1}{30}, \pm 1\right)$ &
                $\left[
                  (3,2,\frac{1}{6})(1,\frac{5}{3})
                  + (\overline{3},1,-\frac{2}{3})(1,-\frac{29}{18})
                  + (\overline{3},1,\frac{1}{3})(1,-\frac{31}{18})
                  + (1,2,-\half)(1,-5)
                \right.$
                \\&&&
                $\left.
                  + (1,1,1)(1,\frac{89}{18})
                  + (1,1,0)(1,\frac{91}{18})
                \right] + \left\{
                  (1,2,\pm\half)(-2,\mp\frac{1}{18})
                  + (1,1,0)(4,0)
                \right\}$
                \\&&&
                  $
                  + (\overline{3},1,\frac{1}{3})(-2,\frac{10}{3})
                  + (3,1,-\frac{1}{3})(-2,-\frac{10}{3})
                 $
                \\\midrule
                (4b.i)&$\checkmark$ &$ \left(\frac{1}{10},-\frac{1}{30},\pm 1\right)$ &
                $\left[
                  (\overline{3},2,-\frac{1}{6})(-1,-5)
                  + (3,1,\frac{2}{3})(-1,\frac{29}{6})
                  + (3,1,-\frac{1}{3})(-1,\frac{31}{6})
                \right.
                $
                \\&&&
                $\left.
                  + (1,2,\half)(-1,15)
                  + (1,1,-1)(-1,-\frac{89}{6})
                  + (1,1,0)(-1,-\frac{91}{6})
                \right]$
                \\&&&
                  $ + \left\{
                  (1,2,\mp\half)(2,\pm\frac{1}{6})
                  + (1,1,0)(-4,0)
                  \right\}
                  + (3,1,-\frac{1}{3})(2,-10)
                  + (\overline{3},1,\frac{1}{3})(2,10)
                 $
                \\&\\
                 (4b.ii)&$\times$&$\left(-\frac{1}{10},-\frac{2}{15},0\right)$ &
                 $
                  ( \overline{3},2,-\frac{1}{6})(-1,0)
                  + (3,1,\frac{2}{3})(-1,0)
                  + (1,1,-1)(-1,0)
                  + (1,2,-\frac{1}{2})(2,0)
                 $
                 \\&&&
                 $
                  + (1,2,\frac{1}{2})(\half,1)
                  + (1,2,\frac{1}{2})(\half,- 1)
                  + (3,1,-\frac{1}{3})(\half,1)
                  + (3,1,-\frac{1}{3})(\half,-1)
                 $
                 \\&&&
                 $
                  + (1,1,0)(-\frac{5}{2}, 1)
                  + (1,1,0)(-\frac{5}{2},-1)
                  + (\overline{3},1,1/3)(2,0)
                 $\\
        \midrule
                (4c.i)&$\times$ &$\left(-\half,\pm \frac{1}{10},\pm \frac{1}{10}\right)$&
                $
                    (\overline{3},2,-\frac{1}{6})(2,\frac{1}{3})
                  + (\overline{3},1,\frac{2}{3})(3,\frac{73}{6})
                  + (1,1,-1)(1,-\frac{23}{2})
                  + (1,2,-\half)(-5,-\frac{25}{2})
                $
                \\&&&
                $
                  + (1,2,\half)(2,-1)
                  + (1,2,\half)(-3,\frac{25}{2})
                  + (3,1,-\frac{1}{3})(1,-\frac{77}{6})
                $
                \\&&&
                $
                  + (3,1,-\frac{1}{3})(-4,\frac{2}{3})
                  + (1,1,0)(8,0)
                  + (1,1,0)(3,\frac{27}{2})
                  + (\overline{3},1,\frac{1}{3})(-4,-\frac{2}{3})
                  $
                \\&\\
                (4c.ii)&$\times$ &$\left(\frac{1}{4},-\frac{3}{20},\frac{1}{10}\right)$&
                $
                  (\overline{3},2,-\frac{1}{6})(\frac{1}{2},\frac{44}{45})
                  + (3,1,\frac{2}{3})(-1,-\frac{62}{45})
                  + (1,1,-1)(2,\frac{10}{3})
                  + (1,1,-\half)(\half,\frac{2}{5})
                $
                \\&&&
                $
                  + (1,1,\half)(-\frac{15}{2},-\frac{5}{3})
                  + (1,1,\half)(\frac{11}{2},-\frac{5}{3})
                  + (3,1,-\frac{1}{3})(7,\frac{31}{45})
                $\\
                &&&
                $
                    + (3,1,-\frac{1}{3})(-6,\frac{31}{45})
                    + (1,1,0)(7,\frac{19}{15})
                    + (1,1,0)(-6,\frac{19}{15})
                    + (\overline{3},1,\frac{1}{3})(-1,-\frac{88}{45})
                  $\\
        \bottomrule
        \end{tabular}
        \caption{\textit{\small The $E_6$ fundamental of restriction to $\gsm\times U(1)_{y'}\times
                U(1)_{y''}$ for all the different chains and all the possible rotations
                reproducing the hypercharge.  In column (2) we show the rotation from the
                $(x_1,x_2,x_3)$ basis to the $(y,y',y'')$ basis that has been used. See text
                for details.}
                \label{tab:tab2}}
        \end{table}
        \begin{table}[htpb]
            \footnotesize
          \centering
        \begin{tabular}{lrrrrrr}
        \toprule
                ID & 1.i & 2.i & 2.ii & 3.i & 4a.ii & 4b.i \\
                $\kappa$& 18 & 3/2 & 21/16 & 4 & 90 & -90\\
        \bottomrule
        \end{tabular}
        \caption{\textit{\small The $\kappa$ values for the chains that admit \pd.}
                \label{tab:tab3}}
        \end{table}

        In \cref{tab:tab2}, we write down the fundamental of $E_6$ on
        restriction to $\gsm\times U(1)_{y'}\times U(1))_{y''}$ for all the
        chains. For all the chains, the hypercharge is given by the linear
        combination of $(x_1,x_2,x_3)$, as defined by $(C_1, C_2, C_3)$, given
        in \cref{tab:tab1}. We have not explicitly shown the breaking $U(1)\to
        Z_N$ for the chains, as that should be clear from the discussion in
        \cref{sec:symmbrk}, specifically \cref{eq:m_par,eq:nn,eq:mm}. We do,
        however, mark each entry by a $\checkmark$ or $\times$ to denote whether
        it can or cannot incorporate a DM stabilising \pd. We have defined $y'$
        and $y''$ in such a way that it is always $y'$ that leaves a remnant
        symmetry, leading us to identifying it as $U(1)_D$ in
        \cref{sec:symmbrk}. There is no chain where both the broken directions
        leave a discrete remnant. From the table, we note that for all the
        chains the remnant $Z_N$ is $Z_2$. For the chains that preserve \pd, we
        put the SM fermions within brackets and the Higgsinos (Higgses for the
        scalar sector) within braces. From \cref{tab:tab2}, we find that for the
        allowed chains, the multiplets transform in the same way under
        $U(1)_{D}$, as pointed out in \cref{sec:symmbrk}. Also, under
        $U(1)_{y''}\equiv U(1)'$, the SM fermions cancel anomalies among
        themselves and the dark-sector fermions are all vectorial. Once the
        $U(1)'$ charges are appropriately scaled, they can be written in the
        form given in \cref{tab:part}. We indicate the $\kappa$ values for the
        different chains in \cref{tab:tab3}. To get the $\kappa$ values, we need
        to scale the $U(1)'$ charges to make the charges of the doublet
        Higgsinos $\pm 1$. It should be pointed out that in organising the
        seemingly different charge assignments as seen in \cref{tab:tab2}, we
        have essentially mapped all the cases that allow \pd\ to the
        $SO(10)\times U(1)$ route. The fundamental of $E_6$, on restriction to
        $SO(10)\times U(1)$ is:
        \begin{align}
          27\to 16(1) + 10(-2) + 1(4),
        \end{align}
        as can be seen from the second block of \cref{tab:tab1}. The
        SM fermions reside in the $SO(10)$ 16, with the Higgsinos in the 10 and
        the 1, the heavy quarks populating the rest of the 10. The $U(1)$
        charges are then exactly the same as $U(1)_D$. That is, for all the
        allowed chains, the multiplets transform under $U(1)_D$ as they would
        for the $SO(10)$ case.

        Although we do not go into tedious discussions on all the chains, we
        would like to present a few clarifications. In row 3ii, we have
        $y',y''=0$ for $h_u$. Therefore, we need to look at the charges of $h_d$
        to determine the remnant $Z_N$. Now, $h_d$ breaks $y'$ down to identity
        for all the multiplets. However, $y''$ breaks to  $Z_3$ for the colour
        triplets and anti-triplets. This $Z_3$ is not a new symmetry on top of
        $\gsm$, as it is isomorphic to the discrete centre of the SM $SU(3)_c$.
        On the point of triplets, we note that in the chains 4b and 4c, the
        $SU(2)$ doublet quark is an $SU(3)$ anti-triplet, while the $SU(2)$
        singlet quarks are triplets. This indicates that for these chains, it is
        the $E_6$ anti-fundamental that gives the particle content with the
        fundamental containing the anti-particles. Accordingly, as shown in
        \cref{tab:tab2}, the spectrum, as obtained from the 27, has the opposite
        hypercharges for these chains.

        At the end of \cref{sec:symmbrk}, while discussing the effects of the
        vev of the scalar giving Majorana mass to the RH neutrinos, we allude to
        a scenario where both the SM doublets are neutral under the $U(1)$ and
        hence play no part in the determination of the remnant $Z_N$. As is
        clear from \cref{tab:tab2}, this scenario does not arise in any of the
        $E_6$ chains. However, for the sake of completeness, we do comment on
        it. For $x_{h_u}=0$, the Dirac mass term for the neutrinos imply
        $x_{\nu^c}= -x_\ell$ and the Majorana mass term implies $x_{\delta_R}=
        -2 x_{\nu^c}= -2 x_\ell$. Therefore, the scalar must have twice the
        charge of the SM leptons. Careful use of the anomaly cancellation
        conditions (see, e.g., \cite{Bandyopadhyay:2018cwu}) and the invariance
        of the Yukawas shows that $\delta_R$ has a charge that is an integral
        ($n\geq 2$) multiple of the charges of all the fermions. This is, for
        example, what happens for the $B-L$ extension of the SM. There, the
        remnant $Z_N$ is completely determined by the charges of the fermions
        and that of the $U(1)_{B-L}$ breaking scalar.

        On that note, we make a final remark before ending the discussion on
        discrete symmetries. In $B-L$ extensions of the SM, the $U(1)_{B-L}$ is
        orthogonal to the $U(1)_Y$, with the SM Higgs being neutral under it. In
        that case, it is perfectly fine to determine the $Z_2$ transformation
        properties of the SM particle content and any exotic multiplets by the
        ratio of $B-L$ charges of the multiplet and that of the $B-L$ breaking
        scalar. However, one often sees this practice carried over to DM studies
        in left-right symmetric (LR) models. This practice might have persisted
        because the bi-doublet in LR models that breaks the electroweak symmetry
        is neutral under $B-L$. However, in LR models, $B-L$ is not orthogonal
        to hypercharge, and it is a combination of $B-L$ and the diagonal
        generator of $SU(2)_R$ that gives hypercharge, the orthogonal direction
        being broken. This is the case depicted in row 3i of \cref{tab:tab2}.
        Note that the up- and down-type scalars are charged under this
        orthogonal direction that spontaneously breaks. Hence, as discussed in
        this text, one needs to look at the $Z_2$ that remains after the SM
        Higgs doublets get a vev. As we see from row 3i, the SM Higgs doublets
        do preserve a remnant $Z_2$, and everything works out. The reason that
        it works out is that the modulus of the charges of $h_u$ and $h_d$ are
        the same. It is slightly misleading to come to this conclusion from the
        $B-L$ assignments alone, as it does not take into account the charges of
        the multiplets under the diagonal of $SU(2)_R$.

    \section{\texorpdfstring{Decompositions of representations of the
    different $E_6$ groups}{Decompositions of representations of the
    different E6 groups}\label{sec:irreps}}%
        In this appendix, we quote a few of the relevant decompositions and direct
        products of the symmetry groups in question. A full list can be found in
        \cite{Slansky:1981yr}.
        \begin{table}
          \footnotesize
        \begin{tabular}{lcl}
        \toprule
        \toprule
        SU(3) &$\to$& SU(2)${\times}$U(1)\\
        {3} & = & $({1})(-2)+({2})(1)$\\
        {6} & = & $({1})(-4)+({2})(-1)+({3})(2)$\\
        {8} & = & $({1})(0)+({2})(3)+({2})(-3)+({3})(0)$\\
        \midrule
        SU(4)& $\to$ &SU(3)${\times}$U(1)\\
         {4} & = & $( {1})(-3)+( {3})(1)$\\
         {6} & = & $( {3})(-2)+( \bar{3})(2)$\\
         {10} & = & $( {1})(-6)+( {3})(-2)+( {6})(2)$\\
         {15} & = & $( {1})(0)+( {3})(4)+( \bar{3})(-4)+( {8})(0)$\\
         \midrule
        SU(5)& $\to$ &SU(4)${\times}$U(1)\\
         {5} & = & $( {1})(-4)+( {4})(1)$\\
         {10} & = & $( {4})(-3)+( {6})(2)$\\
         {15} & = & $( {1})(-8)+( {4})(-3)+( {10})(2)$\\
         {24} & = & $( {1})(0)+( {4})(5)+( \bar{4})(-5)+( {15})(0)$\\
         \midrule
        SU(5)& $\to$ &SU(3)${\times}$SU(2)\\
         {5} & = & $( {1}, {2})(-3)+( {3}, {1})(2)$\\
         {10} & = & $( {1}, {1})(-6)+( \bar{3}, {1})(4)+( {3}, {2})(-1)$\\
         {15} & = & $( {1}, {3})(-6)+( {3}, {2})(-1)+( {6}, {1})(4)$\\
         {24} & = & $( {1}, {1})(0)+( {1}, {3})(0)+( {3}, {2})(5)$
            \\&& $+( \bar{3}, {2})(-5)+( {8}, {1})(0)$\\
         \midrule
        SU(6)& $\to$ &SU(5)${\times}$U(1)\\
         {6} & = & $( {1})(-5)+( {5})(1)$\\
         {15} & = & $( {5})(-4)+( {10})(2)$\\
         {20} & = & $( {10})(-3)+( \bar{10})(3)$\\
         {21} & = & $( {1})(-10)+( {5})(-4)+( {15})(2)$\\
         {35} & = & $( {1})(0)+( {5})(6)+( \bar{5})(-6)$\\
         && +$( {24})(0)$\\
         \midrule
         SU(6)& $\to$ &SU(4)${\times}$SU(2)${\times}$U(1)\\
         {6} & = & $( {1}, {2})(-2)+( {4}, {1})(1)$\\
         {15} & = & $( {1}, {1})(-4)+( {4}, {2})(-1)+( {6}, {1})(2)$\\
         {20} & = & $( {4}, {1})(-3)+( \bar{4}, {1})(3)+( {6}, {2})(0)$\\
         {21} & = & $( {1}, {3})(-4)+( {4}, {2})(-1)+( {10}, {1})(2)$\\
         {35} & = & $( {1}, {1})(0)+( {1}, {3})(0)+( {4}, {2})(3)$
            \\&&
         +$( \bar{4}, {2})(-3)+( {15}, {1})(0)$\\
         \midrule
        SU(6)& $\to$ &SU(3)${\times}$SU(3)${\times}$U(1)\\
         {6} & = & $( {3}, {1})(1)+( {1}, {3})(-1)$\\
         {15} & = & $( \bar{3}, {1})(2)+( {1}, \bar{3})(-2)+( {3}, {3})(0)$\\
         {20} & = & $( {1}, {1})(3)+( {1}, {1})(-3)+( {3}, \bar{3})(-1)$
            \\&&$+( \bar{3}, {3})(1)$\\
         {21} & = & $( {3}, {3})(0)+( {6}, {1})(2)+( {1}, {6})(-2)$\\
         {35} & = & $( {1}, {1})(0)+( {3}, \bar{3})(2)+(\bar{3}, {3})(-2)$
          \\
          &&$+( {8}, {1})(0)+( {1}, {8})(0)$\\
          \bottomrule
          \bottomrule
         \end{tabular}\qquad
         \begin{tabular}{rcl}
         \toprule
         \toprule
        SO(10)& $\to$ &SU(5)${\times}$U(1)\\
         {10} & = & ${5}(2)+ \overline{5}(-2)$\\
         {16} & = & ${1}(-5)+ \overline{5}(3)+( {10})(-1)$\\
         {45} & = & ${1}(0)+ {10}(4)+ \overline{10}(-4)+ {24}(0)$\\
         {54} & = & ${15}(4)+\overline{15}(-4)+{24}(0)$\\
         {120} & = & $ {5}(2)+\overline{5}(-2)+{10}(-6)$
             \\&&$+\overline{10}(6)+{45}(2)+\overline{45}(-2)$\\
         {126} & = & ${1}(10)+{5}(2)+\overline{10}(6)$
             \\&&$+{15}(-6)+\overline{45}(-2)+ {50}(2)$\\
        \midrule
        SO(10) & $\to$ &SU(2)${\times}$SU(2)${\times}$SU(4)\\
         {10} & = & $( {2}, {2}, {1})+( {1}, {1}, {6})$\\
         {16} & = & $( {2}, {1}, {4})+( {1}, {2}, \overline{4})$\\
         {45} & = & $( {3}, {1}, {1})+( {1}, {3}, {1})+( {2}, {2}, {6})$\\
            &&$+( {1}, {1}, {15})$\\
         {54} & = & $( {1}, {1}, {1})+( {3}, {3}, {1})+( {2}, {2}, {6})$\\
          &&$+( {1}, {1}, {20})$\\
         {120} & = & $( {2}, {2}, {1})+( {3}, {1}, {6})+( {1}, {3}, {6})$
         \\&&$+( {1}, {1}, {10})+( {1}, {1}, \overline{10})+( {2}, {2}, {15})$\\
         {126} & = & $( {1}, {1}, {6})+( {3}, {1}, \overline{10})+( {1}, {3}, {10})$\\
          &&$+( {2}, {2}, {15})$\\
        \midrule
        $E_6$ & $\to$ &SU(3)${\times}$SU(3)${\times}$SU(3)\\
        $27$ &=& $(1,\overline{3},3)+(3,3,1)+(\overline{3},\overline{3},1)$ \\
        $78$ &=& $(1,8,1)+(1,1,8)+(8,1,1)$\\
          &&$+(\overline{3},3,3)+(3,\overline{3},\overline{3})$\\
        $351'$ &=& $(1,\overline{3},3) + (\overline{3},1,\overline{3}) + (3,3,1)   $\\
         && $+(1,6,\overline{6})+(\overline{6},\overline{6},1)+(6,1,6)$ \\
          &&$ + (3,3,8) + (\overline{3},8,\overline{3}) + (8,\overline{3},3)$\\%
          $351$ &=& $(1,\overline{3},3) + (\overline{3},1,\overline{3}) + (3,3,1)+(1,\overline{3},\overline{6})$\\
          &&$+(\overline{3},1,6)+(\overline{6},3,1)+(6,1,\overline{3})+(1,6,3)$\\
          &&$+(3,\overline{6},1)+(3,3,8)+(\overline{3},8,\overline{3})+ (8,\overline{3},3)$\\
        \midrule
        $E_6$ & $\to$ & $SO(10){\times}U(1)$\\
        $27$ &=& $1(4)+10(-2)+16(1)$\\
        $78$ &=& $1(0)+45(0)+16(-3)+\overline{16}(3)$\\
        $351'$ &=& $1(-8) + 10(-2) + \overline{16}(-5) + 54(4)$\\
              && $+ \overline{126}(-2) + 144(1)$\\
          $351$ &=& $10(-2)+\overline{16}(-5)+16(1)+45(4)$
          \\&&$+120(-2)+144(1)$\\
        \midrule
        $E_6$ & $\to$ & $SU(6){\times}SU(2)$\\
        $27$ &=& $(\overline{6},2)+(15,1)$\\
        $78$ &=& $(1,3)+(35,1)+(20,2)$\\
        $351'$ &=& $(15,1)+(21,3)+(\overline{84},2)+(105',1)$\\
          $351$&=& $(\overline{6},2)+(21,1)+(15,3)$\\&&
          $+(105,1)+(\overline{84},2)$\\
         \bottomrule
         \bottomrule
        \end{tabular}
        \caption{\textit{\small Irreps of different groups on restriction to subgroups.} \label{tab:irrep}}
        \end{table}

        \begin{table}[htpb]
          \footnotesize
          \centering
         \begin{tabular}{lclclcl}
        $E_6$& & & & SO(10)& & \\
         \toprule
         $27\times \overline{27}$ & = & $1+78+650$&&                                      $16\times 16$ & = & $10 + 120 + 126$ \\
         $27\times 27$ & = & $\overline{27} + \overline{351'} + \overline{351}$&&         $16\times \overline{16}$ & = & $1 + 45 + 210$ \\
         $27\times 78$ & = & $27 + 351 + 1728$&&                                          $16\times 10$ & = & $\overline{16} + \overline{144}$ \\
         $27\times 351'$ & = & $\overline{27} + \overline{1728} + \overline{7722}$&&     $16\times \overline{126}$ & = & $\overline{16} + \overline{560} + \overline{1440}$ \\
         \bottomrule
        \end{tabular}
        \caption{\textit{\small Direct products of the relevant irreps of $E_6$ and $SO(10)$.}
        \label{tab:prod}}
        \end{table}
\end{appendix}

\printbibliography
\end{document}